\documentclass[12pt,epsf]{article}

\usepackage{times}
\usepackage{graphicx}
\usepackage{amsfonts}
\usepackage{amsmath}
\usepackage{amssymb}
\usepackage{amstext}
\usepackage{amsthm}

\usepackage{epsfig}
\usepackage{color}
\psfigdriver{dvips}
\usepackage{latexsym}
\usepackage[matrix,curve]{xypic}
\usepackage{rotating}
\rotdriver{dvips}

\begin{document}

\baselineskip 6mm
\renewcommand{\thefootnote}{\fnsymbol{footnote}}


\newcommand{\nc}{\newcommand}
\newcommand{\rnc}{\renewcommand}


\rnc{\baselinestretch}{1.24}    
\setlength{\jot}{6pt}       
\rnc{\arraystretch}{1.24}   

\makeatletter
\rnc{\theequation}{\thesection.\arabic{equation}}
\@addtoreset{equation}{section}
\makeatother



\nc{\be}{\begin{equation}}

\nc{\ee}{\end{equation}}

\nc{\bea}{\begin{eqnarray}}

\nc{\eea}{\end{eqnarray}}

\nc{\xx}{\nonumber\\}

\nc{\ct}{\cite}

\nc{\la}{\label}

\nc{\eq}[1]{(\ref{#1})}

\nc{\newcaption}[1]{\centerline{\parbox{6in}{\caption{#1}}}}

\nc{\fig}[3]{

\begin{figure}
\centerline{\epsfxsize=#1\epsfbox{#2.eps}}
\newcaption{#3. \label{#2}}
\end{figure}
}


\def\CA{{\cal A}}
\def\CC{{\cal C}}
\def\CD{{\cal D}}
\def\CE{{\cal E}}
\def\CF{{\cal F}}
\def\CG{{\cal G}}
\def\CH{{\cal H}}
\def\CK{{\cal K}}
\def\CL{{\cal L}}
\def\CM{{\cal M}}
\def\CN{{\cal N}}
\def\CO{{\cal O}}
\def\CP{{\cal P}}
\def\CR{{\cal R}}
\def\CS{{\cal S}}
\def\CU{{\cal U}}
\def\CV{{\cal V}}
\def\CW{{\cal W}}
\def\CY{{\cal Y}}
\def\CZ{{\cal Z}}


\def\IB{{\hbox{{\rm I}\kern-.2em\hbox{\rm B}}}}
\def\IC{\,\,{\hbox{{\rm I}\kern-.50em\hbox{\bf C}}}}
\def\ID{{\hbox{{\rm I}\kern-.2em\hbox{\rm D}}}}
\def\IF{{\hbox{{\rm I}\kern-.2em\hbox{\rm F}}}}
\def\IH{{\hbox{{\rm I}\kern-.2em\hbox{\rm H}}}}
\def\IN{{\hbox{{\rm I}\kern-.2em\hbox{\rm N}}}}
\def\IP{{\hbox{{\rm I}\kern-.2em\hbox{\rm P}}}}
\def\IR{{\hbox{{\rm I}\kern-.2em\hbox{\rm R}}}}
\def\IZ{{\hbox{{\rm Z}\kern-.4em\hbox{\rm Z}}}}


\def\a{\alpha}
\def\b{\beta}
\def\d{\delta}
\def\ep{\epsilon}
\def\ga{\gamma}
\def\k{\kappa}
\def\l{\lambda}
\def\s{\sigma}
\def\t{\theta}
\def\w{\omega}
\def\G{\Gamma}


\def\half{\frac{1}{2}}
\def\dint#1#2{\int\limits_{#1}^{#2}}
\def\goto{\rightarrow}
\def\para{\parallel}
\def\brac#1{\langle #1 \rangle}
\def\curl{\nabla\times}
\def\div{\nabla\cdot}
\def\p{\partial}


\def\Tr{{\rm Tr}\,}
\def\det{{\rm det}}


\def\vare{\varepsilon}
\def\zbar{\bar{z}}
\def\wbar{\bar{w}}
\def\what#1{\widehat{#1}}


\def\ad{\dot{a}}
\def\bd{\dot{b}}
\def\cd{\dot{c}}
\def\dd{\dot{d}}
\def\so{SO(4)}
\def\bfr{{\bf R}}
\def\bfc{{\bf C}}
\def\bfz{{\bf Z}}

\begin{titlepage}


\hfill\parbox{3.7cm} {{\tt arXiv:2309.02056}}

\vspace{5mm}

\begin{center}

{\Large \bf  Generalization of Instanton-Induced Inflation \\ and Dynamical Compactification}

\vspace{7mm}

Jeongwon Ho ${}^{a}$\footnote{freejwho@gmail.com}, Kyung Kiu Kim ${}^{b}$\footnote{kimkyungkiu@kookmin.ac.kr},
Seoktae Koh ${}^{c,d}$\footnote{kundol.koh@jejunu.ac.kr}
and Hyun Seok Yang ${}^e$\footnote{hsyang@gist.ac.kr}\footnote{These authors contributed equally to this work.}
\\[10mm]

${}^a$ {\sl Center for Quantum Spacetime, Sogang University, Seoul 121-741, Korea}

${}^b$ {\sl College of General Education, Kookmin University, Seoul 02707, Korea}

${}^c$ {\sl Department of Science Education, Jeju National University, Jeju, 63243, Korea}

${}^d$ {\sl Institute for Gravitation and the Cosmos, The Pennsylvania State University, \\ University Park, PA
16802, USA}

${}^e$ {\sl Department of Physics and Photon Science, Gwangju Institute of Science and Technology, Gwangju 61005, Korea}

\end{center}

\thispagestyle{empty}

\vskip1cm


\centerline{\bf ABSTRACT}
\vskip 4mm
\noindent

It was shown that Yang-Mills instantons on an internal space can trigger the expansion of our four-dimensional universe
as well as the dynamical compactification of the internal space.
We generalize the instanton-induced inflation and dynamical compactification to general Einstein manifolds with positive curvature
and also to the FLRW metric with spatial curvature.
We explicitly construct Yang-Mills instantons on all Einstein manifolds under consideration and
find that the homogeneous and isotropic universe is allowed only if the internal space is homogeneous.
We then consider the FLRW metric with spatial curvature as a solution of the eight-dimensional Einstein-Yang-Mills theory.
We find that open universe $(k=-1)$ admits bouncing solutions unlike the other cases $(k=0, +1)$.
\\



\vspace{0.5cm}

\today

\end{titlepage}

\renewcommand{\thefootnote}{\arabic{footnote}}
\setcounter{footnote}{0}

\section{Introduction}

To understand the origin of the universe, we must combine the general relativity with quantum theory,
dubbed as quantum gravity.
String theory is a leading candidate for the fundamental theory of quantum gravity
which requires a six-dimensional internal space in addition to our four-dimensional spacetime  \cite{st-book1,st-book2}.
However, most quantum gravity theories also require unification of forces in which
gravity plays an equivalent role with gauge theories.
If so, both gravity and gauge theory should play an important role in the early evolution of the universe
such as the cosmic inflation \cite{inflation1,inflation2,inflation3,inflation4}
(see also reviews \cite{kolb-turner} and \cite{Baumann:2014nda}).
In particular, this implies that nonperturbative objects in gravity and gauge theories may play an important role
for the inflationary epoch of the early universe. These are gravitational instantons in gravity such as
Einstein manifolds \cite{gh-cmp79} and Yang-Mills instantons in gauge theories \cite{bpst}.
See also reviews \cite{besse}, \cite{egh-report} and \cite{rajaraman}.

Instantons are defined as topologically nontrivial and nonsingular solutions of classical equations such as
Yang-Mills equations or Einstein equations that minimize the action functional within
their topological type \cite{egh-report}. Mathematically, an instanton is a self-dual or anti-self-dual curvature
in a vector bundle over a four-dimensional Riemannian manifold \cite{don-kro}.
A nonperturbative effect will be important in the strong coupling regime.
Because gravity is also a gauge theory like Yang-Mills theory, gravitational instantons also satisfy the same kind of self-duality equations as Yang-Mills instantons. Indeed it was shown \cite{yang-col1,oh-yang,yang-col2,yang-col3,yang-col4} that
Einstein manifolds can be understood as Yang-Mills instantons for the local Lorentz group $SO(4)$ or $Spin(4)$.
An important point is that Yang-Mills instantons formed in extra dimensions provide a quantized energy density
in our four-dimensional spacetime \cite{kky}.
Therefore, in order to test the role of instantons in the inflationary epoch of the early universe,
one may consider a four-dimensional internal space on which Yang-Mills instantons are supported.
Following the motivation in \cite{kky}, we will study the eight-dimensional Einstein-Yang-Mills theory
in a more general setup by considering the FLRW universe with spatial curvature and general compact Einstein manifolds.
We explicitly construct Yang-Mills instantons on the internal space that generate
the energy-momentum in our four-dimensional spacetime.

In order for both string theory and inflation to be correct, the internal space has to be compactified with
a microscopic size and only four-dimensional spacetime remains in a macroscopic scale
after the end of cosmic inflation. This picture suggests that the cosmic inflation of our four-dimensional universe
would be closely related to the dynamical compactification of extra dimensions.
A recent study has found a model realizing such behavior by showing that the cosmic expansion of our four-dimensional
spacetime causes a dynamical compactification of extra dimensions simultaneously \cite{kky}.
In this paper we will clarify why such behavior must be generic for the eight-dimensional Einstein-Yang-Mills theory.

An accelerating expansion from extra dimensions and a dynamical compactification of extra dimensions are an old idea
explored in many literatures \cite{inf-exd1}-\cite{dyn-comp8} (see also chapter 11.4 in \cite{kolb-turner}) and similar ideas using
monopoles and instantons in extra dimensions have appeared in \cite{old-refs1}-\cite{my-friends2}.
However, their interconnections have not been significantly explored.

This paper is organized as follows. In section 2, we generalize the instanton-induced inflation and dynamical compactification
in \cite{kky} to a more general eight-dimensional spacetime $\mathcal{M}_8$ where the Lorentzian part of $\mathcal{M}_8$ includes a spatial curvature and the four-dimensional internal space is a general Einstein manifold
with positive curvature. We explicitly construct Yang-Mills instantons on all Einstein manifolds under consideration.
We find that the homogeneous and isotropic universe is possible only when the internal space is homogeneous,
so a nonhomogeneous Einstein manifold must be excluded. We then consider the FLRW metric
with spatial curvature as a solution of the eight-dimensional Einstein-Yang-Mills theory.
We find that open universe $(k=-1)$ admits bouncing solutions unlike the other cases $(k=0, +1)$.
In section 3, we discuss how our instanton-induced inflation and dynamical compactification
can evade no-go theorems in \cite{no-go1,swampland2}.
We also comment on some more generalizations and possible implications of the instanton-induced inflation,
in particular, the dynamical compactification of extra dimensions and the reheating mechanism in inflationary cosmology.
Appendix A provides useful algebraic relations for the 't Hooft symbols which are crucially used
for the explicit construction of Yang-Mills instantons on a compact Einstein manifold.
In appendix B, we present the explicit solutions of Yang-Mills instantons on all Einstein manifolds under consideration
and confirm that a nonhomogeneous Einstein manifold does not satisfy the cosmological principle.

\section{Generalization of instanton-induced inflation and dynamical compactification}

We will generalize the basic idea in \cite{kky} by considering a more general eight-dimensional spacetime $\mathcal{M}_8$ where the Lorentzian part of $\mathcal{M}_8$ includes a spatial curvature and the four-dimensional internal space is a general Einstein manifold
with positive curvature.
Consider an eight-dimensional spacetime $\mathcal{M}_8$ whose metric is given by
\begin{eqnarray}\label{prod-metric}
    ds^2 &=& G_{MN} dX^M dX^N = e^A \otimes e^A \xx
    &=& g_{\mu\nu} (x) dx^\mu dx^\nu + e^{2f(x)} h_{\alpha\beta} (y) dy^\alpha dy^\beta
    = e^m \otimes e^m + e^a \otimes e^a,
\end{eqnarray}
where $X^M = (x^\mu, y^\alpha), \; M, N = 0, 1, \cdots, 7; \mu,\nu = 0,1,2,3;
\alpha, \beta =4,5,6,7$, are local coordinates on $\mathcal{M}_8$ and $e^A = (e^m, e^a)
\; A, B = 0, 1, \cdots, 7; m,n = 0,1,2,3; a, b =4,5,6,7$, are orthonormal vielbeins in
$T^* \mathcal{M}_8$. The warped product metric \eq{prod-metric} is necessary in order to incorporate the back-reaction
from Yang-Mills instantons on a four-dimensional internal space.
Consider an $SU(2)$-bundle $\pi: E \to \mathcal{M}_8$ over $\mathcal{M}_8$ whose curvature is given by
\begin{eqnarray} \label{g-curvature}
  F &=& dA + A \wedge A = \frac{1}{2} F_{MN} (X) dX^M \wedge dX^N \\
    &=& \frac{1}{2} \Big( \partial_M A_N - \partial_N A_M + [A_M, A_N]  \Big) dX^M \wedge dX^N
\end{eqnarray}
where $A = A^i_M (X) \tau^i dX^M = \big( A_\mu (x,y) dx^\mu, A_\alpha (x,y) dy^\alpha \big)$
is a connection one-form of the $SU(2)$-bundle $E$ and
$\tau^i \;\big(i=1,2,3 \big)$ are Lie algebra generators with a normalization $\mathrm{Tr} \tau^i \tau^j = - \delta^{ij}$
and obey the commutation relation
\begin{equation}\label{lie-comm}
    [\tau^i, \tau^j] = - \varepsilon^{ijk} \tau^k.
\end{equation}

In order to investigate the dynamical behavior of the eight-dimensional spacetime $\mathcal{M}_{8}$
induced by Yang-Mills instantons on the internal space $X_4$,
let us consider the eight-dimensional Einstein-Yang-Mills theory with the action
\begin{equation}\label{total-action}
    S_{EYM} = \frac{1}{16 \pi G_8} \int_{\mathcal{M}_8} d^8 X \sqrt{-G} R + \frac{1}{4 g_{YM}^2} \int_{\mathcal{M}_8} d^8 X \sqrt{-G} G^{MP} G^{NQ} \mathrm{Tr} F_{MN} F_{PQ}.
\end{equation}
The gravitational field equations read
\begin{equation}\label{einstein-eq}
    R_{MN} - \frac{1}{2} G_{MN} R = 8 \pi G_8 T_{MN}
\end{equation}
with the energy-momentum tensor given by
\begin{equation}\label{em-tensor}
    T_{MN} = - \frac{1}{g_{YM}^2} \mathrm{Tr} \Big( G^{PQ} F_{MP} F_{NQ}
    - \frac{1}{4} G_{MN} F_{PQ}F^{PQ} \Big).
\end{equation}
The action \eq{total-action} also leads to the equations of motion for Yang-Mills gauge fields
\begin{equation}\label{ym-eom}
    G^{MN} D_M F_{NP} = 0.
\end{equation}

If $X_4$ is an Einstein manifold with a positive cosmological constant, that is, $R^{(0)}_{\alpha\beta} = \lambda h_{\alpha\beta}$ with $\lambda > 0$, then $X_4$ is a closed compact manifold
with $vol(X_4) = \int_{X_4} d^4 y \sqrt{h}$ finite \cite{gh-cmp79}.
We are interested in the gauge field configuration on the Einstein manifold $X_4$ described by
\begin{equation}\label{8-gauge}
    A_\mu (x, y) = 0, \qquad A_\alpha (x,y) = A_\alpha (y),
\end{equation}
for which the Yang-Mills action reduces to
\begin{equation}\label{8-4-action}
    S_{YM} = \frac{1}{4 g_{YM}^2} \int_{\mathcal{M}_{3,1}} d^4 x  \sqrt{-g}
    \int_{X_4} d^4 y \sqrt{h} h^{\alpha\gamma} h^{\beta\delta} \mathrm{Tr} F_{\alpha\beta} F_{\gamma\delta}.
\end{equation}
For the gauge field configuration \eq{8-gauge}, the equations of motion \eq{ym-eom} thus take the simple form
\begin{equation}\label{ym-geom}
 h^{\alpha\beta} D_\alpha F_{\beta\gamma} = 0.
\end{equation}
Suppose that the gauge fields \eq{8-gauge} are Yang-Mills instantons on $X_4$
satisfying the self-duality equation \cite{egh-report,rajaraman}
\begin{equation}\label{self-dual-eq}
  F_{\alpha\beta} = \pm \frac{1}{2} \frac{\varepsilon^{\xi\eta \gamma\delta}}{\sqrt{h}}
  h_{\alpha\xi} h_{\beta\eta} F_{\gamma\delta}.
\end{equation}
Then, Eq. \eq{ym-geom} is automatically satisfied and the Yang-Mills instanton solves the equations of motion \eq{ym-eom} even
in a warped spacetime with the metric \eq{prod-metric}.

An important point is that Yang-Mills instantons on the internal space $X_4$ generate a nontrivial energy-momentum tensor given by
\begin{eqnarray}\label{emtensor-1}
&& T_{\mu\nu} = \frac{1}{4 g_{YM}^2} \widetilde{g}_{\mu\nu} \mathrm{Tr} F_{\alpha\beta} F^{\alpha\beta}, \xx
&& T_{\alpha\beta} = - \frac{e^{-2f(x)}}{g_{YM}^2} \mathrm{Tr}
\Big( h^{\gamma\delta} F_{\alpha\gamma} F_{\beta\delta}
    - \frac{1}{4} h_{\alpha\beta} F_{\gamma\delta}F^{\gamma\delta} \Big) = 0, \\
&&  T_{\mu\alpha} = 0, \nonumber
\end{eqnarray}
where $\widetilde{g}_{\mu\nu} (x) = e^{-4f(x)} g_{\mu\nu} (x)$ and, for convenience, all indices are raised and lowered
with the product metric \eq{prod-metric} with $f(x)=0$.
Let us denote the energy-momentum tensor $T_{\mu\nu}$ as the form
\begin{equation}\label{intanton-emtensor}
T_{\mu\nu} = - \frac{1}{g_{YM}^2} \widetilde{g}_{\mu\nu} (x) \rho_n (y),
\end{equation}
where $\rho_n (y) = - \frac{1}{4} \mathrm{Tr} F_{\alpha\beta} F^{\alpha\beta}$ is the instanton density on $X_4$
which is uniform along the four-dimensional spacetime $\mathcal{M}_{3,1}$.
The coupling in Eq. \eq{intanton-emtensor} implies an interesting behavior of four-dimensional spacetime ${\cal M}_{3,1}$.
The energy-momentum tensor $ T_{\mu\nu}$ depends on the size of internal space $X_4$ characterized by the warp factor $e^{2f(x)}$.
For a fixed instanton density, to be specific, if the internal space becomes small, i.e.,
$f(x)$ decreasing, then the energy-momentum tensor $ T_{\mu\nu}$ becomes large, i.e.,
${\cal M}_{3,1}$ more expanding, or vice versa.
This behavior leads to a dynamical mechanism for the compactification of an internal space through the inflation
of our four-dimensional spacetime. Recently, Ref. \cite{kkt} generalized the eight-dimensional
Einstein-Yang-Mills theory by including coupling with a perfect-fluid matter that admits several interesting solutions
such as bouncing universes and oscillatory solutions converging to a de Sitter spacetime.

Suppose that the solution of an $SU(2)$ Yang-Mills instanton on an Einstein manifold $X_4$
is known. In appendix B, we explicitly construct a Yang-Mills instanton on a general Einstein manifold $X_4$
with positive curvature.
Then, it is enough to solve the Einstein equations \eq{einstein-eq} for the warped product geometry \eq{prod-metric}
which take the form
\begin{eqnarray}\label{eins-i11}
&& R_{\mu\nu}^{(0)} - \frac{1}{2} g_{\mu\nu} R_{(g)}
= 4 \big( \nabla^{(g)}_\mu \partial_\nu f
    + \partial_\mu f\partial_\nu f \big) - \big( 4 \nabla^2_{(g)} f
    + 10 g^{\rho\sigma} \partial_\rho f\partial_\sigma f - 2 e^{-2f} \lambda \big) g_{\mu\nu} \xx
   && \hspace{3.3cm} - \frac{8 \pi G_8}{g_{YM}^2} e^{-4f(x)} \rho_n (y) g_{\mu\nu}, \\
\label{eins-i22}
&& R_{\alpha\beta}^{(0)}  - 2 \lambda h_{\alpha\beta} = 3  e^{2f(x)}
\Big( \frac{1}{6} R_{(g)} - \nabla^2_{(g)} f - 2 g^{\mu\nu} \partial_\mu f\partial_\nu f \Big) h_{\alpha\beta},
\end{eqnarray}
where $R_{\mu\nu}^{(0)}$ and $R_{\alpha\beta}^{(0)}$ are the Ricci tensors when $f=0$ and
$\nabla^{(g)}_\mu$ is a covariant derivative with respect to the metric $g_{\mu\nu}(x)$.
A close inspection on the dependence of $\mathcal{M}_{3,1}$
and $X_4$ in Eq. \eq{eins-i22} shows that a solution exists
only if the internal space must be a space with a constant Ricci scalar, i.e., an Einstein manifold.
With this condition, Eq. \eq{eins-i11} further requires that its consistent solution exists
only if the instanton density $\rho_n (y)$ is constant.
This means that the instantons are uniformly smeared out on the internal space $X_4$.
Note that the internal space $X_4$ in our case is a compact Einstein manifold.
It is known \cite{gh-cmp79} that an Einstein manifold with a positive cosmological constant must
be compact and four examples are known: $\mathbb{S}^4, \, \mathbb{C}P^2, \, \mathbb{S}^2 \times
\mathbb{S}^2$ and Page where the Page \cite{page-metric} is an \textit{inhomogeneous} Einstein manifold
on the nontrivial product as a twisted $\mathbb{S}^2$ bundle over $\mathbb{S}^2$.
We will show that the Page space does not admit a constant instanton density,
so it cannot be a solution of the above equations.
It is interesting to note that a homogeneous and isotropic universe cannot be obtained
from an inhomogeneous internal space such as the Page space.
This result suggests that the uniformity of the internal space is closely related to the cosmological principle
of our universe.

To solve the above equations, let us take the specific ansatz  \cite{kolb-turner}
\begin{equation}\label{time-metric}
    ds^2 = -dt^2 + e^{2h(t)} \left(\frac{dr^2}{1-k r^2} +r^2 d\Omega^2 \right) +  e^{2f(t)}
     h_{\alpha\beta} (y) dy^\alpha dy^\beta,
\end{equation}
where $k$ is normalized as $-1,\; 0$, or 1 and $d\Omega^2$ is the metric of unit sphere $\mathbb{S}^2$.\footnote{In our
metric ansatz \eq{time-metric},
$h(t) = \ln a(t)$ where $a(t)$ is a usual scale factor of the four-dimensional FLRW universe,
so $\dot{h} = \frac{\dot{a}}{a}$ corresponds
to the Hubble parameter. Accordingly, $\ddot{h} = \frac{\ddot{a}}{a}
- \frac{\dot{a}^2}{a^2}$.}
The Einstein equations (\ref{eins-i11}) and (\ref{eins-i22}) for the metric \eq{time-metric}
leads to the following differential equations
\begin{align}\label{4dimk-ee1}
E_1 =&\,\, \dot{h}^2 + 4\dot{f}\dot{h} + 2\dot{f}^2 +\frac{8}{\zeta_c^2} e^{-2f}-\frac{1}{3\zeta_I^2} e^{-4f} + k e^{-2 h} = 0,
 \\   \label{4dimk-ee2}
E_2 =&\,\, 2\ddot{h}+4\ddot{f} + 3\dot{h}^2+8\dot{f} \dot{h}
+ 10\dot{f}^2+ \frac{24}{\zeta_c^2}e^{-2f}-\frac{1}{\zeta_I^2} e^{-4f} + k e^{-2 h} = 0,
 \\    \label{4dimk-ee3}
E_3 =&\,\, \ddot{h}+2\dot{h}^2 + \ddot{f} +3\dot{f} \dot{h} + 2\dot{f}^2 +\frac{4}{\zeta_c^2}e^{-2f} +k e^{-2 h} = 0,
\end{align}
where $\zeta_c^2 \equiv \frac{12}{\lambda}$ and $\zeta_I^2 \equiv \frac{g_{YM}^2}{8 \pi G_8 \rho_n}$. 
Note that new terms proportional to $k$ have been appeared.
In general relativity, the metric component $g_{00}$ is not a dynamical variable but a Lagrange multiplier.
Thus $E_1=0$ derived from the variation $\delta g_{00}$ is not an evolution equation
but a (Hamiltonian) constraint that initial conditions for the evolution equations,
$E_2 = 0$ and $E_3=0$, must satisfy.
Indeed, one can show that $\dot E_1 + \big( 4\dot f(t) + 3\dot h(t) \big) E_1 = \dot h(t) E_2 + 4 \dot f(t) E_3$ \cite{kky}.
Therefore, if the evolution equations are solved with initial conditions
chosen to be consistent with $E_1=0$, the constraint equation will always be satisfied.
As a result, the constraint equation will give us an important information for the time evolution
of an eight-dimensional spacetime $\mathcal{M}_8$.

\begin{figure}
\centering
\setlength{\unitlength}{1.3cm}
\begin{picture}(10,3)(-3,-2)
\put(-1.5,0){\vector(1,0){7}}
\put(5.7,-0.1){$\mathcal{X}=e^{-2f}$}
\put(0,-2.5){\vector(0,1){4}}
\put(0.85,0.3){\vector(1,-1){0.3}}
\put(0.4,0.4){$\mathcal{X}_L$}
\put(4.1,0.3){\vector(-1,-1){0.3}}
\put(4.2,0.35){$\mathcal{X}_R$}
\textcolor[rgb]{0.00,0.00,1.00}{\put(2.5,-2){\line(0,1){3}}}
\put(2.4,1.1){$\mathcal{X}_s = \lambda \zeta_I^2$}
{\line(1,0){0.2}}
{\line(1,0){0.2}}
\put(-0.55,1.7)
{$\mathcal{Y}$}
\textcolor[rgb]{0.98,0.00,0.00}{\qbezier(-1,-2)(2,3)(5,-2)}
\put(-0.7,-1.2){$c$}
\end{picture}
\caption{\label{p-graph} The graph of $\mathcal{Y} = a\mathcal{X}^2 + b\mathcal{X} + c$.}
\end{figure}
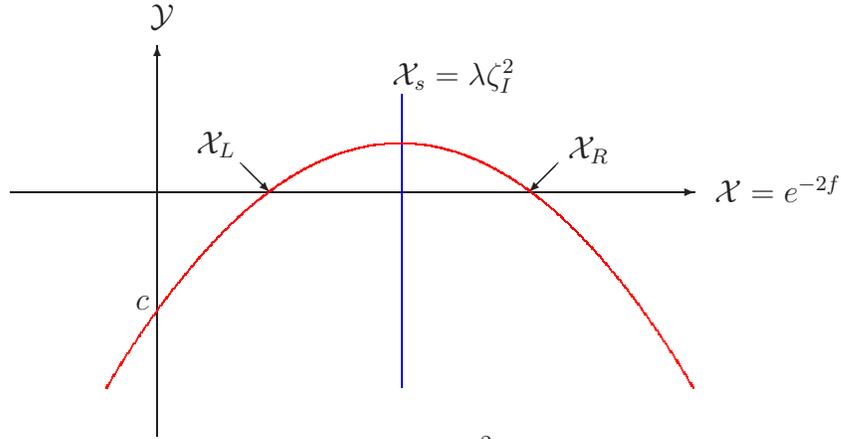

The solution space is most easily understood by studying the constraint equation \eq{4dimk-ee1}.
Let us define a graph for the constraint $\mathcal{Y}: = E_1 = 0$ as
\begin{equation}\label{graph2}
    \mathcal{Y} = a \mathcal{X}^2 + b \mathcal{X} + c = a \Big( \mathcal{X} + \frac{b}{2a} \Big)^2 + \Big(c - \frac{b^2}{4a} \Big),
\end{equation}
where $\mathcal{X} = e^{-2f}, \; a= - \frac{1}{3 \zeta_I^2} < 0, \; b=\frac{2\lambda}{3}\geq0$
and $c = \dot{h}^2 + 2\dot{f}^2  + 4 \dot{f} \dot{h} + k e^{-2 h}$.
The graph \eq{graph2} is thus described by a concave parabola with the symmetric axis
$\mathcal{X}_s = - \frac{b}{2a}= \lambda \zeta_I^2 > 0$.
Then the constraint equation can be understood as
finding $\mathcal{X}$-intercepts for given velocity data $(\dot{h}, \dot{f})$
which determines the $y$-intercept $c$.
One can easily infer the behavior of a solution from Fig. \ref{p-graph}.
The quadratic equation $\mathcal{Y}(\mathcal{X})=0$ has two (in general, complex) roots $\mathcal{X}_L$ and $\mathcal{X}_R$ given by
\begin{equation}\label{x-lr}
    \mathcal{X}_{L} = - \frac{b}{2a} \left(1 - \sqrt{1-\frac{4ac}{b^2}} \right),
    \qquad \mathcal{X}_{R} = - \frac{b}{2a} \left(1 + \sqrt{1-\frac{4ac}{b^2}} \right).
\end{equation}
From our definition $\mathcal{X} = e^{-2f} > 0$, these $\mathcal{X}$-intercepts must be positive.
The symmetric axis $\mathcal{X}_s > 0$ is positive and
solely determined by the ratio $12 \frac{\zeta_I^2}{\zeta_c^2}$.
Hence it does not depend on velocities $(\dot{h}, \dot{f})$.
It means that the parabola in Fig. \ref{p-graph} simply moves up or down along the symmetric axis $\mathcal{X} = \mathcal{X}_s$
depending on velocities $(\dot{h}, \dot{f})$.
Note that the $\mathcal{X}$-intercept $\mathcal{X}_L$ becomes positive only when the $y$-intercept $c$ is negative.

\subsection{Flat and closed universes}

For $k=0$ and $k = +1$ (i.e., flat and closed universes), the condition that $c= \dot{h}^2 + 2\dot{f}^2
+ 4 \dot{f} \dot{h} + k e^{-2 h}$ is negative requires $\dot{f} \dot{h} < 0$ for any solution of the $\mathcal{X}_L$-branch.
Thus there are two possibilities for the solution:
\begin{equation}\label{lr-moving}
  \dot{f} \dot{h} < 0 \quad \Longleftrightarrow   \quad
\left\{
  \begin{array}{ll}
    \dot{f} < 0, \quad \dot{h} > 0, & \hbox{$\mathcal{X}_L$ moving to the right;} \\
    \dot{f} > 0, \quad \dot{h} < 0, & \hbox{$\mathcal{X}_R$ moving to the left.}
  \end{array}
\right.
\end{equation}
The right-moving solution describes an eight-dimensional spacetime where our four-dimensional spacetime is expanding
and the internal space is contracting while the left-moving solution does the opposite.
We will show later that open universe $(k=-1)$ exhibits a completely different behavior;
it admits bouncing solutions unlike the other cases $(k=0, \; +1)$.

A simple argument shows that $f(t)$ in the $\mathcal{X}_L$-branch is a (strictly) monotonic function of $t$.\footnote{The figures
in Ref. \cite{kky} clearly have shown this behaviour.}
Suppose that the function $f(t)$ in the $\mathcal{X}_L$-branch is not monotonic.
It means that the function $f(t)$ can pass through the region where $\dot{f}=0$ at some time.
Since the $\mathcal{X}_L$-branch does not allow $c>0$, $\dot{h} = 0$ there and so $c = 0$ (i.e., $\mathcal{X}_L = 0$).
Therefore $\dot{f}=0$ in the $\mathcal{X}_L$-branch also enforces $\dot{h}=0$, in which the eight-dimensional spacetime
becomes $\mathcal{M}_8 = \mathbb{R}^{3,1} \times X_4$ with $vol(X_4) \to \infty$.
The evolution equations, $E_2 = E_3 = 0$, then imply that this spacetime must continue thereafter.
This behavior is an exceptional case possible only in a particular contracting solution but cannot appear in generic cases,
especially in expanding solutions. Therefore, the warp factor $f(t)$ in the $\mathcal{X}_L$-branch must be a (strictly)
monotonic function of $t$ for generic solutions.
It leads to an interesting picture \cite{inf-exd2} that the cosmic expansion of our four-dimensional spacetime necessarily
accompanies a dynamical compactification of extra dimensions.
However, the $\mathcal{X}_R$-branch admits a solution satisfying $\dot{f} \dot{h} > 0$,
so it does not necessarily require a similar behavior as the $\mathcal{X}_L$-branch.
Now we will argue that the $\mathcal{X}_R$-branch corresponds to a quantum branch
so that it should be discarded within a classical approximation.

A notable point is that the size of internal space depends on the warp factor $e^{2f(t)} = \mathcal{X}^{-1}$
whose value is determined by solving the constraint equation $\mathcal{Y}=E_1=0$.
First, note that the graph in Fig. \ref{p-graph} is a concave parabola,
so the parabola cannot move down below the $\mathcal{X}$-axis since the parabola below the $\mathcal{X}$-axis means
that there is no real solution of Eq. \eq{graph2}.
This implies that the internal space has a minimum size for an $\mathcal{X}_L$-branch solution
while a maximum size for an $\mathcal{X}_R$-branch solution, whose size is set by the value $\mathcal{X}_s$
of the symmetric axis of the parabola. In order to estimate the explicit value of $\mathcal{X}_s$,
let us use the relation, $G_8 = g_{YM}^2 l_s^2$, well-known in string theory (see \S 13.1.3 in \cite{st-book1} and
\S 13.3 in \cite{st-book2}) where $l_s$ characterizes a fundamental scale as the size of string.
Using the explicit result $\rho_1 = \frac{3}{4R^4}$ for the single instanton on $\mathbb{S}^4$,
we get\footnote{\label{diff-ggt}There was a minor error in Eq. (3.19) in \cite{kky}. Considering the difference
between $SU(2)$ generators in gravity and gauge theory, it should read as $\rho_1 (y) = \frac{12}{\zeta_c^4}
= \frac{3}{4 R^4}$. See the footnote 2 in \cite{yang-col4} for the explanation of this difference.}
\begin{equation}\label{r-minimum}
    \mathcal{X}_s = \frac{R^2}{2 \pi l_s^2},
\end{equation}
where $R = \frac{\zeta_c}{2}$ is the radius of the internal space $\mathbb{S}^4$.
Since the size of internal space is given by $R e^{f(x)}$,
we see that $(R e^f) \geq \sqrt{2 \pi} l_s \approx 2.5 l_s$ for the $\mathcal{X}_L$-branch whereas
$0 \leq (R e^f) \leq \sqrt{2 \pi} l_s \approx 2.5 l_s$ for the $\mathcal{X}_R$-branch.
In particular, the $\mathcal{X}_R$-branch travels at most $2.5 l_s$ distances.
Moreover, when $\mathcal{X} \to \infty$, i.e. $e^f \to 0$, the $\mathcal{X}_R$-branch hits the spacetime (internal) singularity.
If the fundamental scale $l_s$ is comparable to the Planck length $l_P = 10^{-35}$ m,
the Planck distance travel cannot be described by a classical geometry.
Hence it is reasonable that the $\mathcal{X}_R$-branch can be ignored within the classical approximation.

We emphasize that the existence of the minimum size of the internal space is due to the Yang-Mills
instanton smeared over the compact space $X_4$. Without the instanton,
the graph in Eq. \eq{graph2} is replaced by a linear equation, i.e., $\mathcal{Y} = b\mathcal{X} + c$
with the same $c= \dot{h}^2 + 2\dot{f}^2  + 4 \dot{f} \dot{h} + k e^{-2 h}$ \cite{inf-exd10}.
In this case, there is only a single branch which can travel the entire positive range, $\mathcal{X}>0$,
but still requires the condition \eq{lr-moving} to have a real root.
Even in this case, $f(t)$ must be a monotonic function of $t$ since $\dot{f} = 0$ enforces $c > 0$ for generic cases,
where $\mathcal{X} = e^{-2f}$ becomes negative, physically not allowed.
The monotonicity of the function $f(t)$ implies that there is no bounce.
A similar behavior is also expected for the case with  Yang-Mills instantons.
Actually the same reasoning implies that the bounce $\mathcal{X}_L (\dot{f} < 0) \rightleftarrows \mathcal{X}_L (\dot{f} > 0)$
is not allowed either for any solution of the differential equations \eq{4dimk-ee1}-\eq{4dimk-ee3}.
However, this situation will change if the four-dimensional space is open universe with $k=-1$.

\subsection{Open universe}

\begin{figure}[hbtp]
\centering
\includegraphics[width=0.4\textwidth]{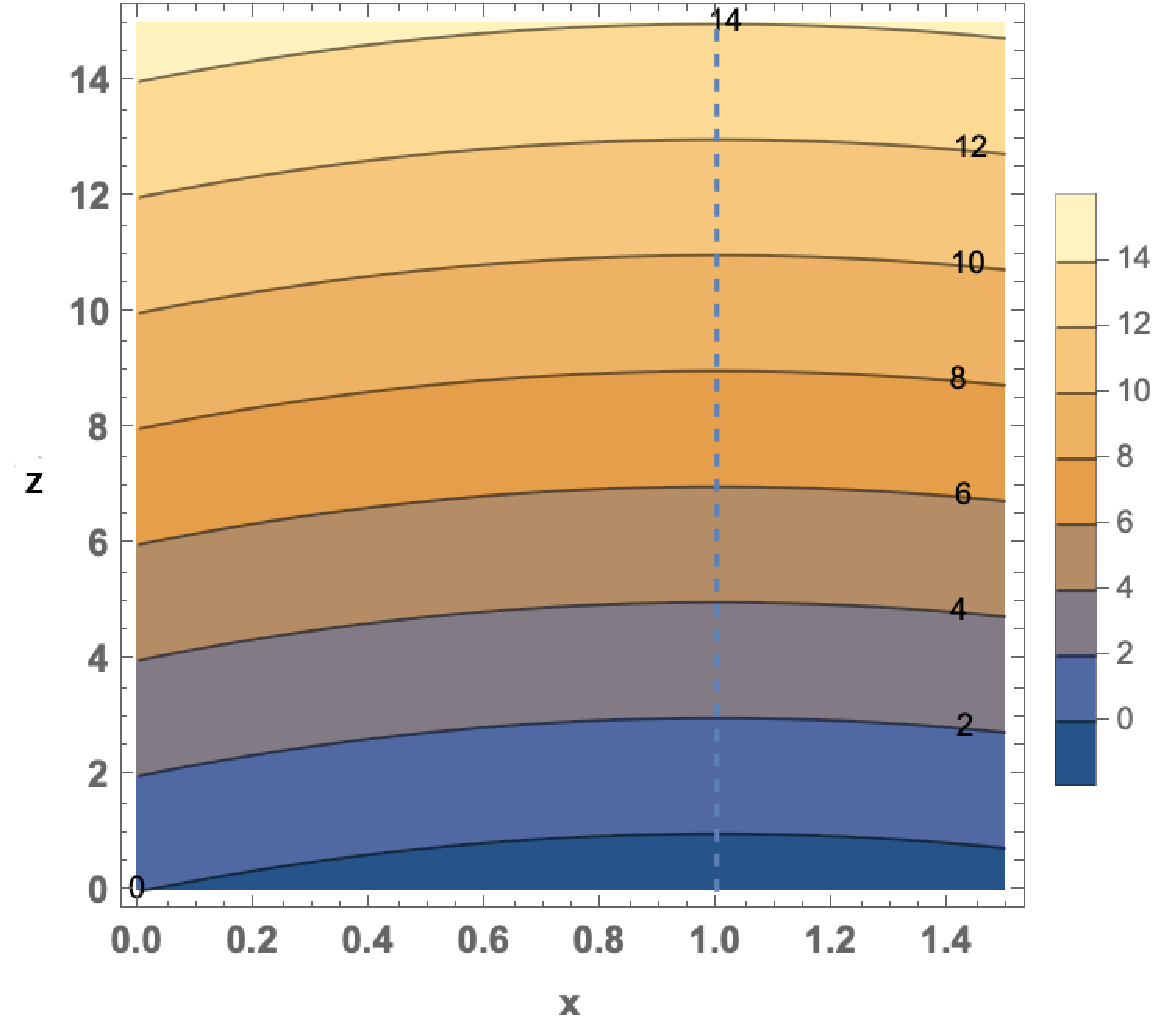}
\includegraphics[width=0.4\textwidth]{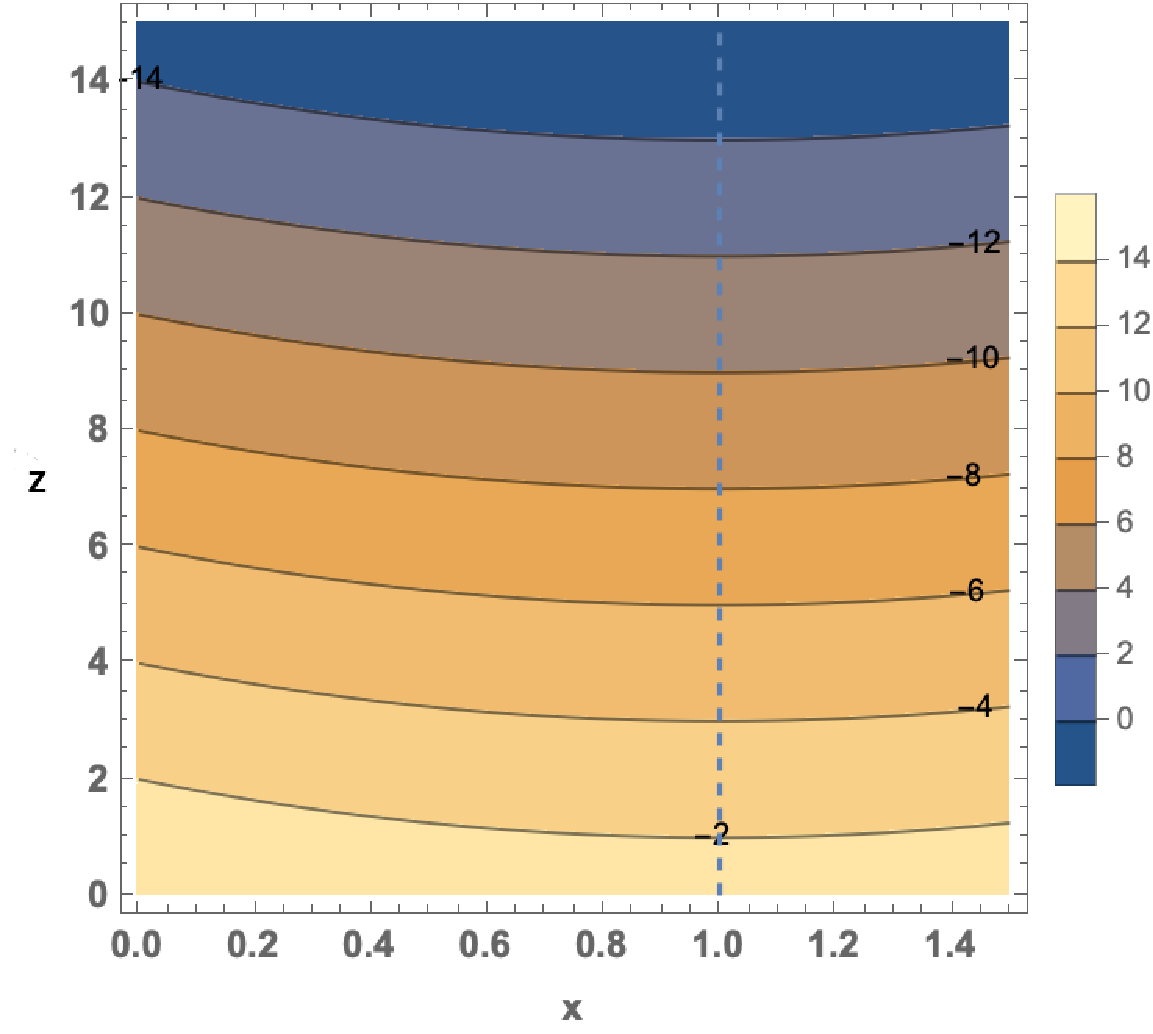}
\caption{\label{fig_nonflat} Contour plot of the constraint equation (\ref{4dimk-ree1}) for $k=-1$ (left) and $k=1$ (right)
in $(\mathcal{X} = e^{-2f}, \, \mathcal{Z}=e^{-2h})$-plane.
Each contour represents $c_0 = 4\dot{f}\dot{h} +2\dot{f}^2 +\dot{h}^2$.}
\end{figure}

For a numerical analysis, it is convenient to rescale the variables in the constraint equation (\ref{4dimk-ee1})
by absorbing the model parameters as
\begin{align}\label{rescale}
t \rightarrow \frac{\zeta_c^2}{4 \sqrt{3} \zeta_I} t, \qquad f \rightarrow f+\log \left(\frac{\zeta_c}{2 \sqrt{3} \zeta_I}\right),
\qquad h \rightarrow h+\frac{1}{2} \log \left(\frac{\zeta_c^4}{48 \zeta_I^2}\right),
\end{align}
The constraint equation (\ref{4dimk-ee1}) can then be expressed with the rescaled parameters as
\begin{align}
    \mathcal{Y}(\mathcal{X}, \mathcal{Z}) \equiv
    4 \dot{f} \dot{h}+2 \dot{f}^2+\dot{h}^2 -\mathcal{X}^2+ 2\mathcal{X} + k\mathcal{Z}=0,
    \label{4dimk-ree1}
\end{align}
where $\mathcal{X} \equiv e^{-2f} > 0$ and $\mathcal{Z} \equiv e^{-2h} > 0$.
The contour plot in the first quadrant of $(\mathcal{X}, \mathcal{Z})$-plane is shown in Fig. \ref{fig_nonflat}
and each contour line represents  $c_0 \equiv 4\dot{f}\dot{h} +2\dot{f}^2 +\dot{h}^2$.
If we solve Eq. (\ref{4dimk-ree1}) in terms of $\mathcal{X}$, the roots are given by
\begin{align}
\mathcal{X}_{ \pm}=1 \mp \sqrt{1 + c_0 + k \mathcal{Z}}.
\label{xpm}
\end{align}
The effect of introducing a new term, $k\mathcal{Z}$, is only to shift the $\mathcal{Y}$-intercept in Fig. \ref{p-graph}
from $c_0$ to $c \equiv c_0 + k\mathcal{Z}$, as shown in Fig. \ref{fig_nonflat},
but not to change the position of the symmetry axis $\mathcal{X}_s = 1$ (or $\lambda \zeta_I^2$).
Therefore $\mathcal{X}_{-}$ corresponds to a quantum branch for the same reason as the $k=0$ case,
so we discard the $\mathcal{X}_{-}$-branch.
Solutions in the $\mathcal{X}_{+}$-branch are constrained by $0 \leq \mathcal{X}_+ \leq 1$ as seen from Eq. (\ref{xpm}).

\begin{figure}[hbtp]
\centering
\includegraphics[width=0.3\textwidth]{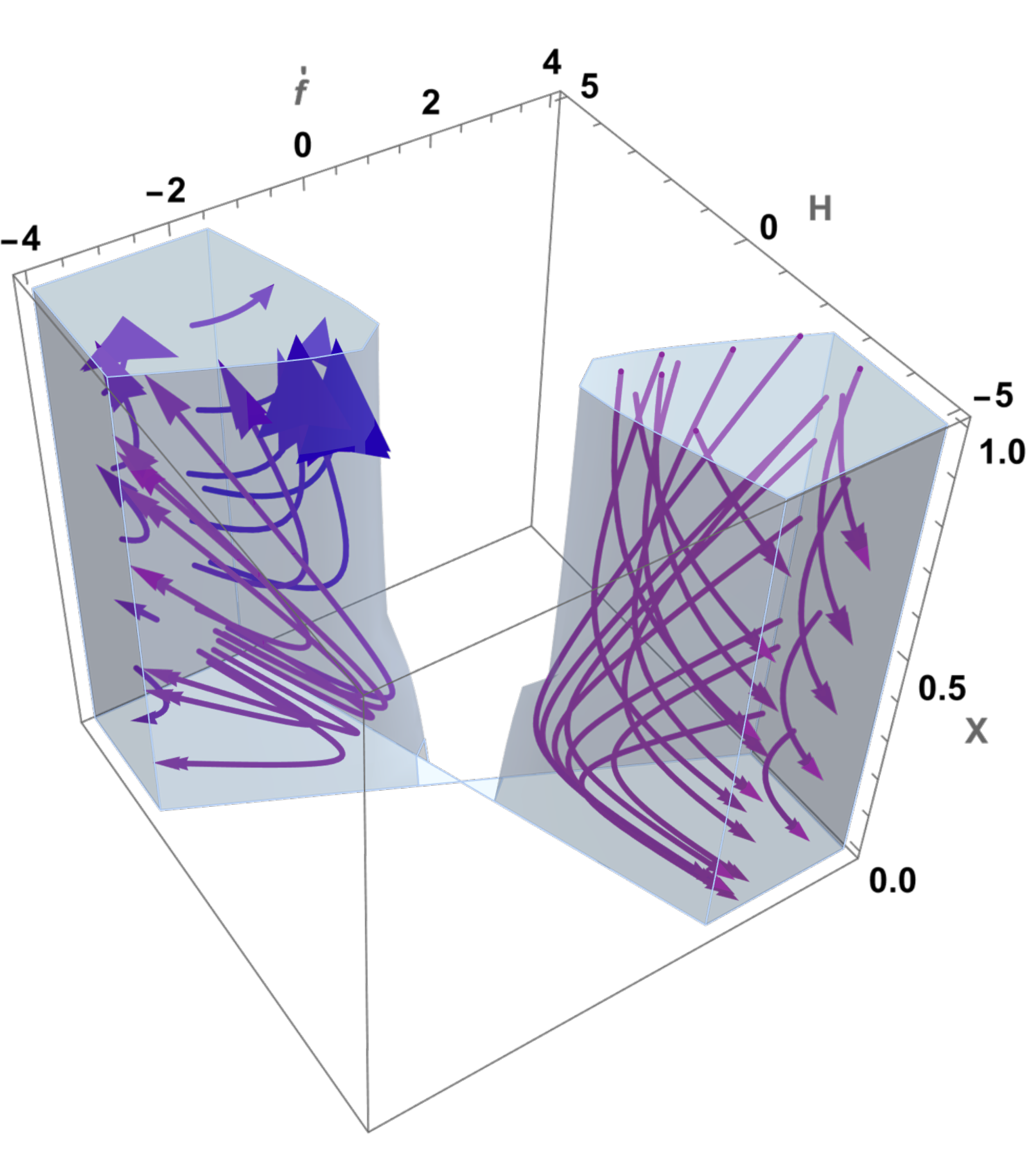}
\includegraphics[width=0.3\textwidth]{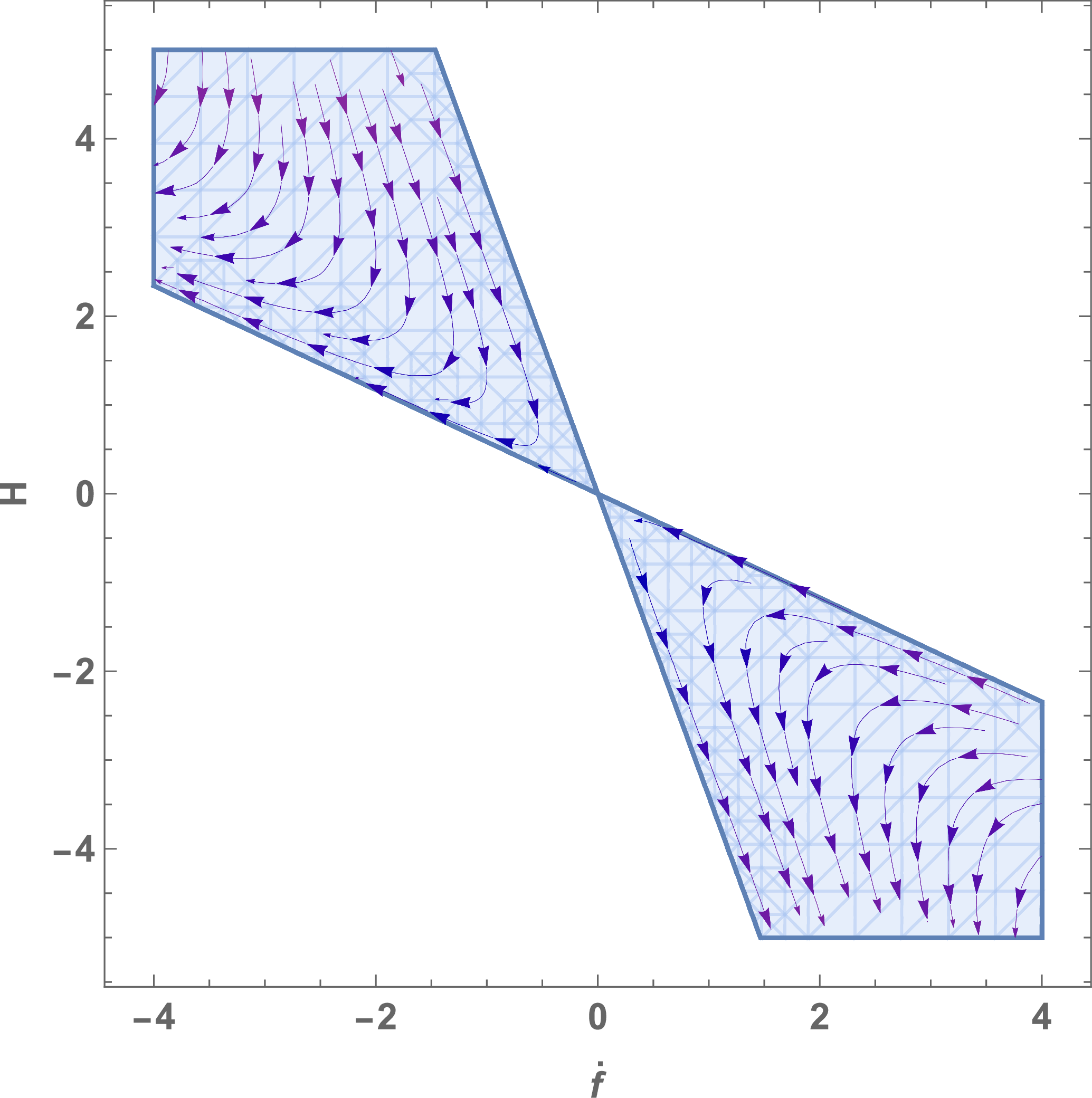}
\includegraphics[width=0.3\textwidth]{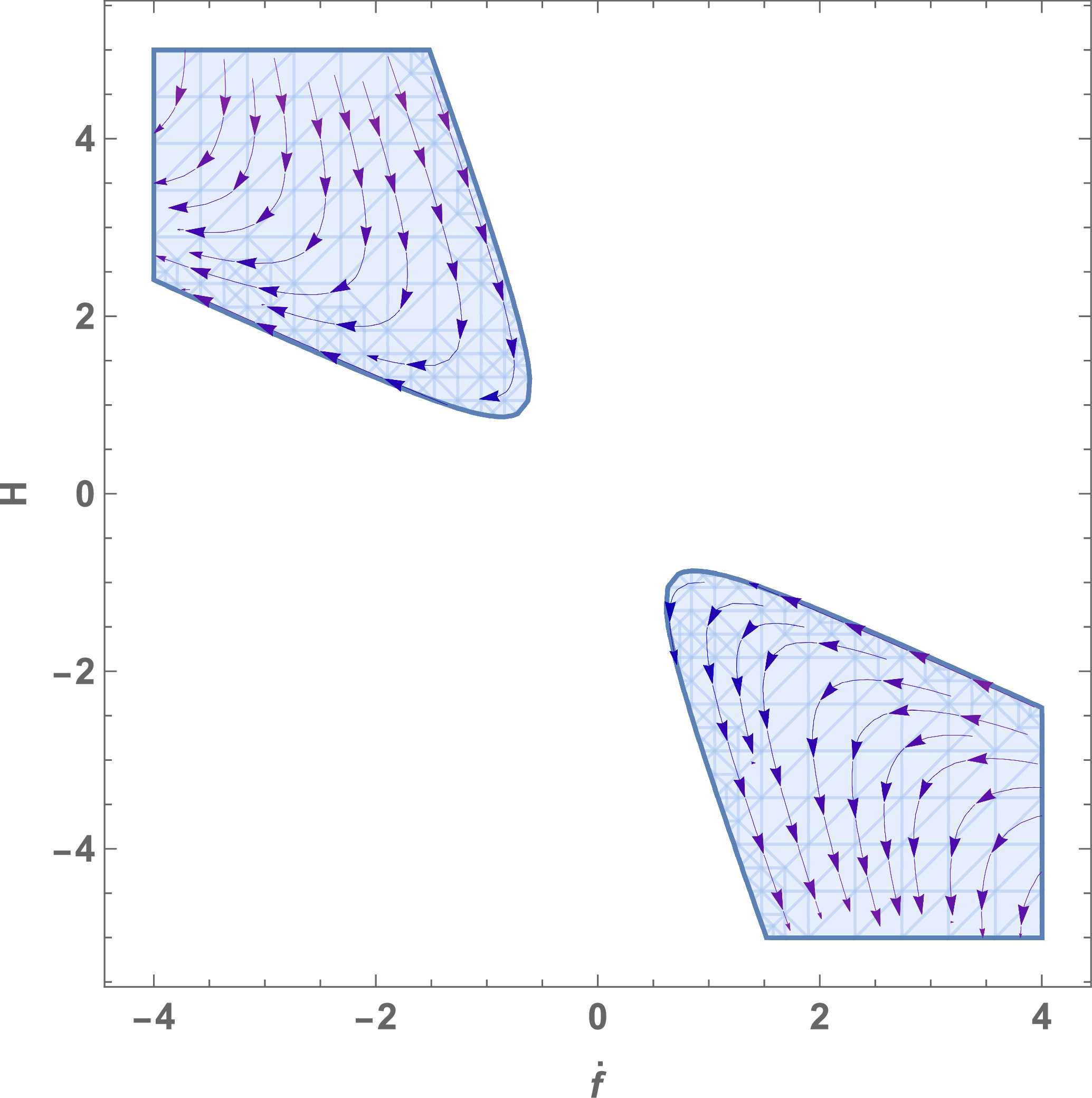}
\caption{\label{k1flow} Solution flows with $k=1$. The middle and right figures show the projection onto constants $X=0$ and $X = 0.5$, respectively. The size of arrow head in the left figure denotes the rapidity of time evolution.}
\end{figure}

\begin{figure}[hbtp]
\centering
\includegraphics[width=0.3\textwidth]{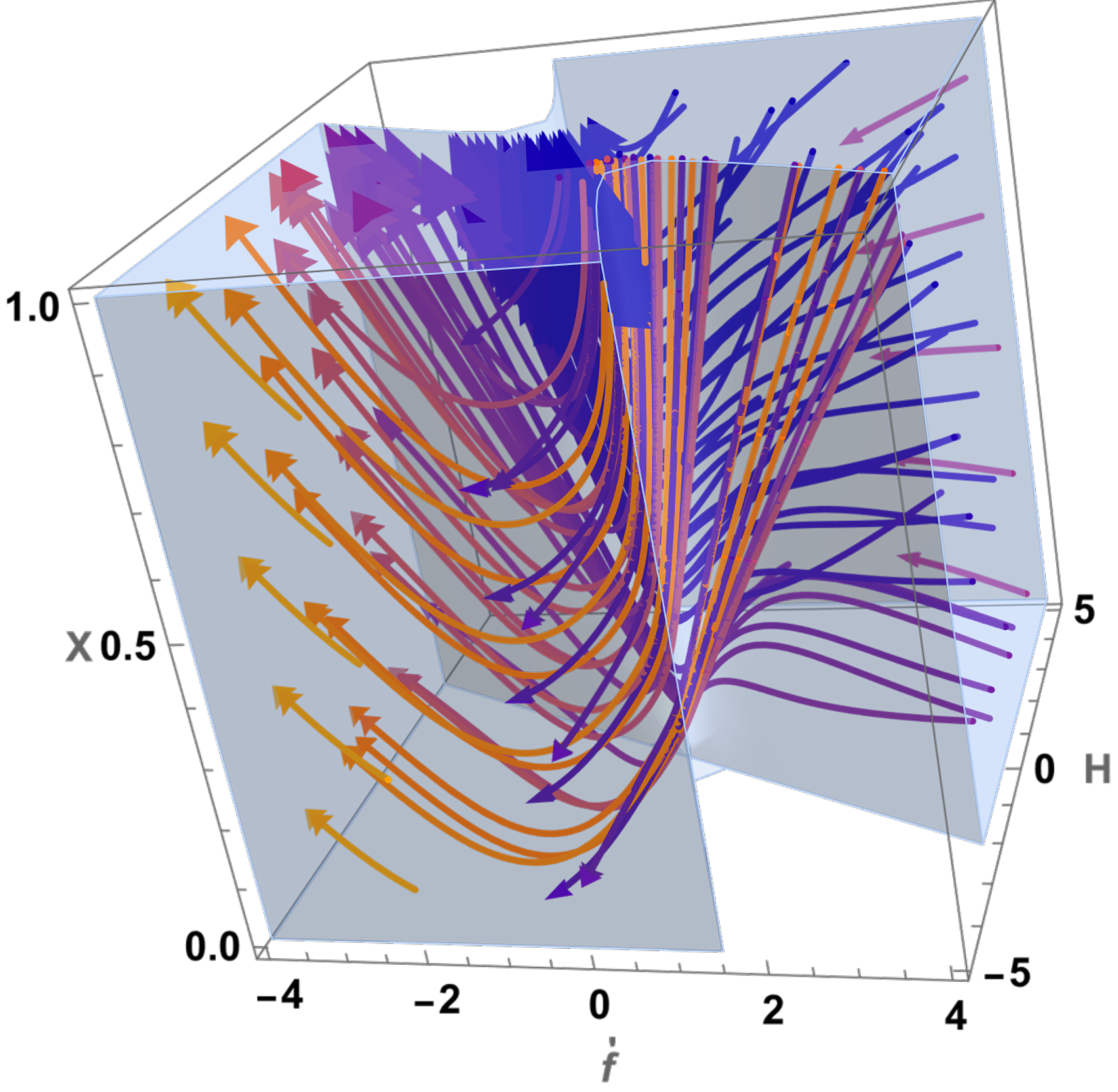}
\includegraphics[width=0.3\textwidth]{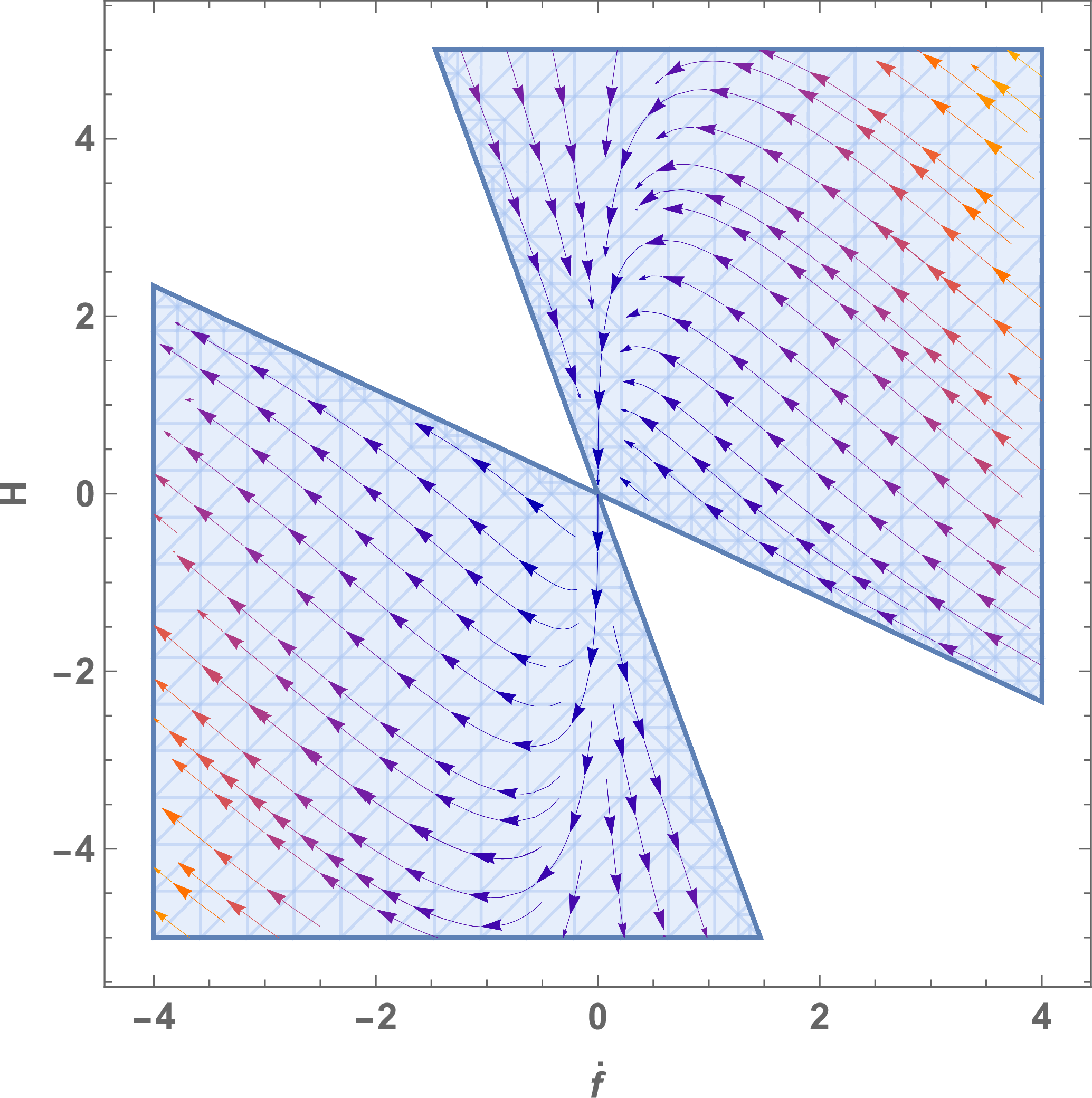}
\includegraphics[width=0.3\textwidth]{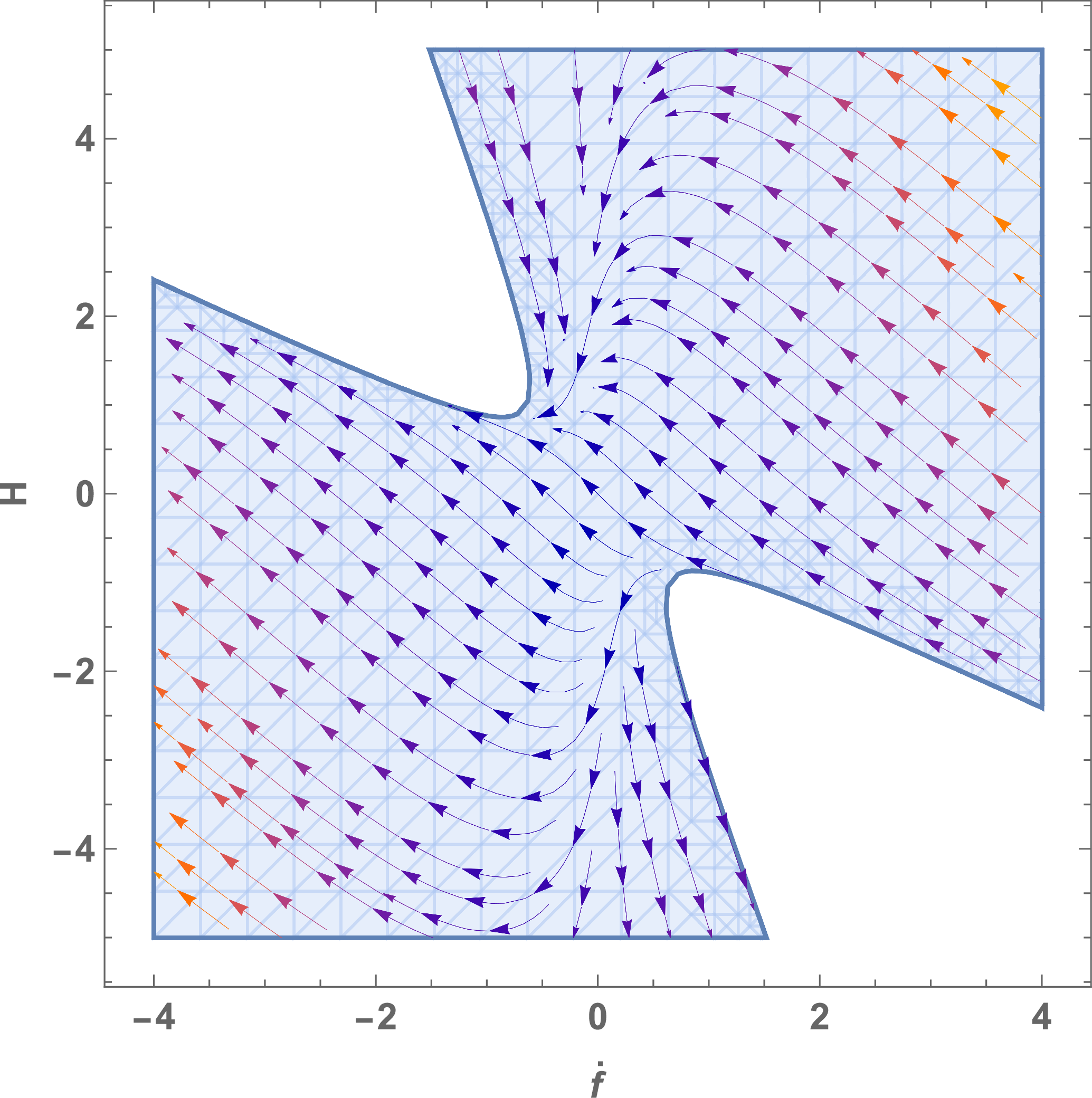}
\caption{\label{km1flow} Solution flows with $k=-1$. The middle and right figures show the projection onto constants $X=0$
and $X = 0.5$, respectively. These flows contain bouncing solutions.}
\end{figure}

Let us rewrite Eqs. (\ref{4dimk-ee1})-(\ref{4dimk-ee3}) as the form
\begin{eqnarray} \label{4dimk-flowx}
&& \ddot{f}= -3 \dot{f} H-4 \dot{f}^2+2 \mathcal{X}^2-3 \mathcal{X}, \xx
&& \dot{H}=4 \dot{f} H+4 \dot{f}^2-H^2-3 \mathcal{X}^2+4 \mathcal{X},\\
&& \dot{\mathcal{X}}= - 2 \dot{f} \mathcal{X}, \nonumber
\end{eqnarray}
where $H \equiv \dot{h}$ is the Hubble parameter and $\mathcal{Z}$ is eliminated using the constraint equation (\ref{4dimk-ree1}),
\begin{align} \label{cons-num}
k \mathcal{Z} = \mathcal{X}^2 -2 \mathcal{X} -4 \dot{f} H-2 \dot{f}^2-H^2.
\end{align}
This set of equations generates three-dimensional flows depending on the data $(\dot{f}, H, \mathcal{X})$.
Given the initial condition $\left(\dot{f},H,\mathcal{X}\right)\big|_{t=t_i}$, these flows show the full-time evolution
of metric functions. The numerical analysis is separately done for $k=+1$ and $k=-1$ cases by solving the constraint \eq{cons-num}
for each case. Figures \ref{k1flow} and \ref{km1flow} have been obtained by solving numerically the flow equations \eq{4dimk-flowx}.

It may be instructive to compare the behavior of solutions in the $\mathcal{X}_{+}$-branch for $k=-1$ case and other cases $(k=0,1)$.
As was pointed out above, the $\mathcal{Y}$-intercept in the graph defined by Eq. \eq{4dimk-ree1}
for a given $\mathcal{Z}$ is given by $c = c_0 + k\mathcal{Z}$, which must be negative to satisfy $\mathcal{X}_+ > 0$.
Therefore, for $k=1$, it is necessary to require $\dot{f} \dot{h} < 0$ in order to obey $c_+ \equiv c_0 + \mathcal{Z} < 0$.
Hence the behavior of solutions for $k=1$ is similar to that for the $k=0$ case.
The three-dimensional solution flows for $k=+1$ are depicted in Fig. \ref{k1flow}.
Because the behavior of solutions with $k=+1$ is similar to the case with $k=0$,
no bounce appears in the $k=+1$ case for the same reason as the $k=0$ case.
However, the $k= -1$ case is very different from the other cases.
Since the $\mathcal{Y}$-intercept in this case is given by $c_- \equiv c_0 - \mathcal{Z}$,
the condition $c_- < 0$ does not necessarily require the condition $\dot{f} \dot{h} < 0$.
So the function $f(t)$ needs not be monotonic.
In other words, a bounce solution is allowed for the $k= -1$ case,
where a solution makes a smooth transition from a region where $\dot{f} > 0$ to a region where $\dot{f} < 0$
or vice virsa.
Fig. \ref{km1flow} clearly shows this behavior.\footnote{This behavior is somewhat counterintuitive.
One might naively expect that a bouncing solution would appear in the $k=1$ case where
the four-dimensional spacetime $\mathcal{M}_{3,1}$ contains a compact space $\mathbb{S}^3$, so back-reactions
from the compact space $\mathbb{S}^3$ would cause a bouncing behavior. But it is not the case.
Rather, the bounce solution appears in the open universe $(k=-1)$. Thus it will be interesting to clarify
a detailed mechanism of the bouncing behavior for the $k= -1$ case.}

Finally, we briefly comment on the four-dimensional Ricci scalar $R(g) = g^{\mu\nu} R_{\mu\nu}$ for the metric (\ref{time-metric}).
Since the $(\mu,\nu)$-components of the eight-dimensional Ricci tensor are given by
\begin{align}
	R_{\mu \nu}=R_{\mu \nu}^{(0)}-4\left(\nabla_\mu^{(g)} \partial_\nu f+\partial_\mu f \partial_\nu f\right),
\end{align}
the Ricci scalar reads
\begin{eqnarray} \label{r-scalar-k}
\tilde{R}(g) &\equiv& \frac{\zeta_c^4}{2^5 3^2 \zeta_I^2} R(g) = -4 \dot{f} \dot{h}- k \mathcal{Z} - \dot{h}^2 - \mathcal{X}^2 \xx
&=& - c + 2 \dot{f}^2  - \mathcal{X}^2,
\end{eqnarray}
after taking into account the rescaling \eq{rescale}.
In order to obtain a de Sitter-like solution, it is necessary to have $\tilde{R}(g) > 0$.
For $k=1$, $c_+ = c_0 + \mathcal{Z} < 0$ requires $\dot{f}\dot{h} < 0$ and the condition $c_+ < 0$ is stronger than
that for the $k=0$ case, $c_0 <0$. Therefore, in order for the four-dimensional FLRW spacetime to have a positive Ricci scalar,
the condition $\dot{f}\dot{h} < 0$ is more stringent than the flat case.
However, the stringent condition is not necessary for the $k=-1$ case since, even if $\dot{f} \dot{h} > 0$,
$\tilde{R}(g)$ could be positive while obeying $c_- = c_0 - \mathcal{Z} < 0$.
As a result, bouncing solutions are also allowed in this case as was illustrated by Fig. \ref{km1flow}.

\subsection{General Einstein manifolds with positive curvature}

The eight-dimensional Einstein-Yang-Mills theory consists of the Einstein equations \eq{einstein-eq} and
the Yang-Mills equations \eq{ym-eom}.
For the ansatz \eq{prod-metric} and \eq{8-gauge}, the Yang-Mills equations \eq{ym-eom} reduce to \eq{ym-geom}
and the Einstein equations \eq{eins-i11} depend only on the instanton density which must be constant
for the existence of a solution.
If one can construct an $SU(2)$ Yang-Mills instanton on a general Einstein manifold $X_4$ that
obeys Eq. \eq{self-dual-eq}, the Yang-Mills equations \eq{ym-geom} are automatically satisfied.
Therefore, it is enough to check that the $SU(2)$ Yang-Mills instanton on an Einstein manifold $X_4$ provides
a constant instanton density $\rho^{(\pm)}_n = - \frac{1}{4} \mathrm{Tr} F^{(\pm)}_{\alpha\beta}F^{(\pm) \alpha\beta}$.
The following formula is useful;
\begin{equation}\label{density}
  F^{(\pm)i} \wedge F^{(\pm)i} = \pm 2 \rho^{(\pm)}_n \sqrt{h} d^4 y
\end{equation}
where $\sqrt{h} d^4 y = e^1 \wedge e^2 \wedge e^3 \wedge e^4$ is the volume form on $X_4$.
If $\rho^{(\pm)}_n$ is constant, the left-hand side of Eq. \eq{density} must be a constant multiple of the volume form.

\begin{table}[h!]
  \begin{center}
    \begin{tabular}{|c|c|c|c|c|}
    \hline
    \multicolumn{1}{|c}{} & \multicolumn{1}{|c} {$\mathbb{S}^4$}  & \multicolumn{1}{|c} {$\mathbb{C}P^2$} &
    \multicolumn{1}{|c} {$\mathbb{S}^2 \times \mathbb{S}^2$} & \multicolumn{1}{|c|} {Page} \\
      \hline
      K\"ahler?  & No  & Yes  & Yes & No  \\ 
      \hline
       Spin?  & Yes  & No  & Yes & No \\
      \hline
      Homogeneous?  & Yes  & Yes  & Yes & No \\
      \hline
    \end{tabular}
    \caption{Compact Einstein manifolds}
    \label{tab-compeins}
  \end{center}
\end{table}

Although the equations of motion \eq{ym-eom} have been reduced to the problem solving the instanton equation \eq{self-dual-eq}
on a compact Einstein manifold $X_4$, it is, in general, difficult to find an instanton solution on the Einstein manifold.
Fortunately there exists a general method \cite{yang-col1,oh-yang} to find an $SU(2)$ Yang-Mills instanton on $X_4$
if the metric on $X_4$ is known. Since the explicit instanton construction is a technical justification of the assumption
already made in \cite{kky}, every details will be delegated to appendices.
Instead, we summarize the obtained results.
In order to realize a homogeneous and isotropic universe, it is necessary to construct
Yang-Mills instantons uniformly smeared out over a compact Einstein manifold $X_4 = \mathbb{S}^4, \; \mathbb{C}P^2,
\; \mathbb{S}^2 \times \mathbb{S}^2$ and Page $=\mathbb{C}P^2 \sharp \overline{\mathbb{C}P^2}$.
We find that such instantons are allowed for all compact Einstein manifolds under consideration except the inhomogeneous Page space.
Table \ref{tab-compeins} provides a summary of the geometric properties of these compact Einstein manifolds,
including the K\"ahler, spin, and homogeneous structures \cite{gh-cmp79,besse}.
The emerging conclusion is that the uniformity of internal space is essential
for realizing a homogeneous and isotropic universe.

\section{Discussion}

There are no-go theorems stating that there are no nonsingular de Sitter compactifications
for large class of gravity theories \cite{no-go1,swampland2}.
We will clarify why our instanton-induced inflation and dynamical compactification can evade such no-go theorems.
Maldacena and Nu\~nez consider a $D$-dimensional warped metric \cite{no-go1}
\begin{equation} \label{MN-ansatz}
ds^2 = \Omega^2(y) \left( g^{(d)}_{\mu\nu}(x) dx^\mu dx^\nu + h^{(D-d)}_{\alpha\beta}(y) dy^\alpha dy^\beta \right).
\end{equation}
Using the $D$-dimensional Einstein equations, they showed that the warping factor should satisfy the equation
\begin{equation} \label{MN-eq}
\frac{d}{(D-2) \Omega^{D-2}} \nabla_{(h)}^2 \Omega^{D-2} = R(g) - \Omega^2 \left(T^\mu_{\;\; \mu} -\frac{d}{D-2} T^M_{\;\;\; M} \right),
\end{equation}
where $R(g)$ is the scalar curvature of the $d$-dimensional metric $g$.
A simple manipulation with Eq. \eq{MN-eq} shows that the de Sitter spacetime with $R(g)>0$
is not allowed without a bare positive cosmological constant.

A main reason for our model to evade the Maldacena-Nu\~nez argument is that the warp factor in Eq. \eq{time-metric}
is time-dependent whereas that in \eq{MN-ansatz} depends only on the coordinates of the internal space.\footnote{Indeed,
the authors in \cite{no-go1} pointed out (see the footnote $i$) that there can be expanding (or contracting)
universe solutions when some scalar fields are time-dependent.}
Let us clarify this aspect for the eight-dimensional Einstein-Yang-Mills theory \eq{total-action}.
The Einstein equations (\ref{einstein-eq}) can be written as the form
\begin{align}
	R_{MN} = 8\pi G_8 \left( T_{MN} - \frac{1}{6} G_{MN} T^{L}_{\;\; L} \right).
\end{align}
Contracting $(\mu,\nu)$-components leads to \eq{r-scalar-k}, i.e.,
\begin{align} \label{r-scalar-ec}
	R(g) = -6 \biggl( \dot{h}^2 +4\dot{f}\dot{h} + k \mathcal{Z} + \frac{1}{3\zeta_I^2} \mathcal{X}^2\biggr).
\end{align}
Since we are interested in a de Sitter-like solution, we will exclude the open universe $(\mathrm{i.e.}, k=-1)$ case.
(In fact, the no-go theorem in Ref. \cite{no-go1} was also restricted to $d$-dimensional metrics
which correspond to Minkowski or de Sitter space.)
If the metric were time-independent, $R(g) = -6 k \mathcal{Z} -\frac{2}{\zeta_I^2} \mathcal{X}^2 < 0$, so
we could not obtain the de Sitter space.
However, if the metric is time-dependent, we can obtain $R(g) > 0$ as long as the term $\dot{f}\dot{h}$ is negative
and large enough. Therefore, in order to have a de Sitter space including expanding universes,
the time-dependence is crucial so that $\dot{f}\dot{h}$ has a negative value.
Thus the dynamical compactification of an internal space is an inevitable consequence of an inflating universe in our toy model.

The no-go theorem in \cite{no-go1} has been generalized in \cite{swampland2} to $D$-dimensional time-dependent metrics
\begin{equation} \label{swamp-metric}
ds^2 = g^{(d)}_{\mu\nu}(x) dx^\mu dx^\nu + e^{2 \rho(x)} h^{(D-d)}_{\alpha\beta}(y) dy^\alpha dy^\beta.
\end{equation}
Starting with the eleven-dimensional supergravity $(D = 11)$, the effective potential in $d$ dimensions
has been obtained by the compactification to $(11-d)$-dimensional curved internal manifolds.
After the Weyl scaling
\begin{equation}\label{weyl-metric}
 g^{(d)}_{\mu\nu}(x) = e^{2 \omega(x)} q^{(d)}_{\mu\nu}(x)
\end{equation}
with
\begin{equation}\label{weyl-scale}
  \omega (x) = - \frac{D-d}{d-2} \rho(x),
\end{equation}
the $d$-dimensional effective theory is defined in the Einstein frame with the potential
\begin{equation}\label{eff-pot}
  V(\chi) = V_R e^{-\lambda_1 \chi} + V_G e^{-\lambda_2 \chi}
\end{equation}
where $\lambda_1 = \frac{6}{\sqrt{(d-2)(11-d)}}, \; \lambda_2 = \frac{2 (d+1)}{\sqrt{(d-2)(11-d)}}$,
and $\chi = 3 \sqrt{\frac{(11-d)}{d-2}} \rho$ is scaled to have a canonically normalized kinetic term of scalar field,
$\frac{1}{2} (\nabla \chi)^2$. In Eq. \eq{eff-pot}, $V_R = - R(h)$ is the scalar curvature of the internal manifold
and $V_G$ (which is always positive) is the average of the four-form $G$-flux in the eleven-dimensional supergravity.
Note that $\lambda_1 \leq \lambda_2$ for $d \geq 2$. Then one can show that the effective potential \eq{eff-pot}
does not admit a de Sitter vacuum $\chi_0$ where $V'(\chi_0)=0$ and $V(\chi_0) >0 $. In particular, for the region where $V > 0$,
$\frac{|\partial_\chi V|}{V}$ is a monotonically decreasing function for both $V_R > 0$ and $V_R < 0$ and
satisfies the swampland bound \cite{swampland2}.

The metric ansatz \eq{swamp-metric} has the same form as ours in \eq{prod-metric}.
We should emphasize that the no-go theorem in \cite{swampland2} is a result derived
from the effective potential \eq{eff-pot}, which is defined in $d$-dimensional Einstein frame with the metric $q_{\mu\nu}$.
However, we did not consider any compactification (or dimensional reduction) to four-dimensional spacetime.
Instead, we have considered the dynamical evolution of the entire eight-dimensional spacetime, rather than
addressing the dynamics in a lower-dimensional effective field theory.
Furthermore, we describe the evolution in the Einstein frame of the eight-dimensional spacetime rather than
the four-dimensional spacetime. As we have noticed, the evolution of the four-dimensional universe and the internal space
is deeply interconnected, so in our case, it is necessary to consider the entire evolution of the eight-dimensional spacetime.
The description based on the four-dimensional effective field theory is not sufficient to capture the dynamical response
of the internal space. As a result, the no-go theorem derived from the lower-dimensional effective potential \eq{eff-pot}
cannot be applied to the eight-dimensional spacetime, where the internal space undergoes dynamical compactification.

It was shown in \cite{swampland2} that the swampland inequality, $\frac{|\nabla V|}{V} \geq \lambda_1$,
related to the no-go theorem discussed above, is a consequence of the fact that the $D$-dimensional theory
satisfies the strong energy condition. It should also hold true in our case.
The strong energy condition states that for any timelike vector $V^M$, $R_{MN} V^M V^N \geq 0$,
or in any local coordinate systems, $R_{00} \geq 0$.
A simple calculation with the metric (\ref{time-metric}) shows that
 \begin{align} \label{R00}
	R_{00} = \frac{1}{2}\biggl(5k \mathcal{Z} + \frac{40}{\zeta_c^2} \mathcal{X} -\frac{1}{\zeta_I^2} \mathcal{X}^2
+ 10\dot{f}^2 +20\dot{f}\dot{h} + 5\dot{h}^2 \biggr),
\end{align}
where Eqs. (\ref{4dimk-ee2}) and (\ref{4dimk-ee3}) are used again.
Substituting (\ref{4dimk-ee1}) into (\ref{R00}) leads to
\begin{align}
	R_{00} = \frac{1}{3\zeta_I^2} \mathcal{X}^2 > 0.
\end{align}
Therefore, we find that the strong energy condition is satisfied in our model for all $k$.
However, it does not imply that a $D$-dimensional gravity theory does not admit a de Sitter-like
solution which describes an expanding or contracting universe, as we discussed above.

Although we have generalized the instanton-induced inflation and dynamical compactification in Ref. \cite{kky}
to general Einstein manifolds with positive curvature and also to the FLRW metric with spatial curvature,
we have omitted (or postponed) an important class of generalization: Turning on Yang-Mills gauge fields
on $\mathcal{M}_{3,1}$ as well as those on $X_4$.
We can consider a more general gauge field configuration
\begin{equation}\label{44-gauge}
    A_\mu (x, y) = A_\mu (x), \qquad A_\alpha (x,y) = A_\alpha (y),
\end{equation}
instead of Eq. \eq{8-gauge}. However, a time-dependent as well as static finite-energy solution
does not exist in pure $SU(2)$ Yang-Mills theory on $\mathbb{R}^{3,1}$ \cite{tdfesym1,tdfesym2}.
Recently it was shown in \cite{ds-yms1,ds-yms2,ds-yms3} that nontrivial solutions with finite energy and finite action
can be constructed in the pure $SU(2)$ Yang–Mills theory if the theory is defined on
four-dimensional de Sitter $dS_4$ and anti-de Sitter $AdS_4$ spaces.
Indeed, if we take $h(t) = \ln \cosh(t)$ with $k=1$ and $h(t) = \ln |\sin(t)|$ with $k= - 1$ in Eq. \eq{time-metric},
the FLRW metric describes the de Sitter space $dS_4$ and the anti-de Sitter space $AdS_4$, respectively.
In particular, since each of the two four-dimensional Yang-Mills theories is conformally
invariant for the ansatz \eq{44-gauge}, it is possible to generalize to any conformally flat spacetime
such as FLRW if one can find isotropic Yang-Mills solutions $A_\mu (x)$ on Minkowski, de Sitter,
or anti-de Sitter spacetime \cite{ds-yms3}.
Therefore it will be interesting to embed the nontrivial solutions in \cite{ds-yms1,ds-yms2,ds-yms3}
into the eight-dimensional Einstein-Yang-Mills theory \eq{total-action} with the ansatz \eq{44-gauge}.
We hope to address this issue in the near future.

Our toy model can be embedded into string theory by introducing a two-dimensional surface $\Sigma$ such that
ten-dimensional spacetime becomes $\mathcal{M}_8 \times \Sigma$
(see \cite{kky} for a brief discussion of such an embedding).
Then it is natural to expect that the inflation of our four-dimensional spacetime
has to simultaneously occur with the dynamical compactification of extra dimensions
because only the four-dimensional spacetime in $\mathcal{M}_8 \times \Sigma$ remains in a macroscopic scale after
the end of cosmic inflation. The internal space must be stabilized in some way with a microscopic size
after the inflation.
We have observed that the internal space in $\mathcal{M}_8$ has a minimum size.
Our model is controlled by two parameters denoted by $\zeta_c^2 \equiv \frac{12}{\lambda}$
and $\zeta_I^2 \equiv \frac{g_{YM}^2}{8 \pi G_8 \rho_n}$.
These are basically determined by the curvature of an internal Einstein manifold $X_4$ and the instanton density on $X_4$,
respectively. Their ratio $\mathcal{X}_s = \frac{12 \zeta_I^2}{\zeta_c^2}$ turns out to be a fundamental scale
which is precisely the minimum size of the internal space. Therefore an interesting question is:
When the extra dimensions approach to this scale, what happens?

For a solution in the $\mathcal{X}_L$-branch moving to the right, the (3+1)-dimensional spacetime
is expanding whereas the four-dimensional internal space is contracting.
Since the internal space $X_4$ has the minimum size at $\mathcal{X} = \mathcal{X}_s$ which is roughly the size of Planck scale
and the warp factor $f(t)$ must be a (strictly) monotonic function for $k= 0, 1$,
the inflation should end near the minimum size while the internal space would be stabilized there.
A crude speculation is that there will be a huge quantum back-reaction from the instanton over
the tiny internal space when the extra dimensions approach to the minimum size.
Then the quantum back-reaction would excite Yang-Mills gauge fields as well as metric fluctuations
in the eight-dimensional Einstein-Yang-Mills theory. It is expected that the inflation energy will be transferred to
ubiquitous fluctuations and rapidly decreased by producing a large amount of radiations and matters.
Ultimately, the cosmic inflation will end and the extra dimensions would be stabilized
around the critical size $\mathcal{X}_s$. If so, the dynamical stabilization of the internal space
may be closely related to the matter generation in our four-dimensional spacetime.
This phenomenon would correspond to a reheating mechanism
in inflationary cosmology \cite{kolb-turner,Baumann:2014nda}.

We showed that the open universe with $k=-1$ (negatively curved) allows a bounce solution.
In four-dimensional Einstein gravity coupled with a scalar field,
it is known \cite{bounce-cosm} that the existence of bouncing solution depends on the spatial curvature.
For a flat or open universe, the possibility of a bounce
is precluded by the null energy condition.
For a closed universe, on the other hand, the bounce can take place when the curvature term balances
with the total energy of the universe. But the bounce solution appeared in the open universe in our case.
There is no violation of the null energy condition.
This might be possible due to an energy exchange between the four-dimensional spacetime and the internal space
and the open universe might be more favorable for the energy exchange.
It should be interesting to clarify a detailed mechanism of the bouncing behavior for the $k= -1$ case.

Another interesting feature in our model is that the homogeneity of internal space is crucial to realize
a homogeneous and isotropic universe.
We found that the Page space \cite{page-metric} does not admit
such universe that satisfies the cosmological principle.
In order to incorporate the Page space within our framework,
it would be necessary to consider a generalized Bianchi-type metric with a warp factor
depending on the extra dimensions too, which may be responsible for an inhomogeneous and
anisotropic universe \cite{non-cos}.

\section*{Acknowledgments}

This work was supported by the National Research Foundation of Korea (NRF)
with grant numbers NRF-2019R1A2C1007396 (KK), NRF-2021R1A2C1005748 (SK) and NRF-2018R1D1A1B0705011314 (HSY).
JH was supported by the Ministry of Education through the Center for Quantum Spacetime (CQUeST) of
Sogang University (NRF-2020R1A6A1A03047877) and NRF-2020R1A2C1014371.
We acknowledge the hospitality at APCTP where part of this work was done.


\appendix

\section{'t Hooft symbols}

The explicit components of the 't Hooft symbols $\eta^i_{ab}$ and ${\overline \eta}^i_{ab}$
for $i = 1,2,3$ are given by
\begin{eqnarray} \label{tHooft-symbol}
\begin{array}{l}
{\eta}^i_{ab} = {\varepsilon}^{i4ab} + \delta^{ia}\delta^{4b}
- \delta^{ib}\delta^{4a}, \\
{\overline \eta}^{i}_{ab} = {\varepsilon}^{i4ab} - \delta^{ia}\delta^{4b}
+ \delta^{ib}\delta^{4a}
\end{array}
\end{eqnarray}
with ${\varepsilon}^{1234} = 1$. They satisfy the following relations \cite{chy-grg}
\begin{eqnarray} \label{self-eta}
&& \eta^{(\pm)i}_{ab} = \pm \frac{1}{2} {\varepsilon_{ab}}^{cd}
\eta^{(\pm)i}_{cd}, \\
\label{proj-eta}
&& \eta^{(\pm)i}_{ab} \eta^{(\pm)i}_{cd} =
\delta_{ac}\delta_{bd}
-\delta_{ad}\delta_{bc} \pm \varepsilon_{abcd}, \\
\label{self-eigen}
&& \varepsilon_{abcd} \eta^{(\pm)i}_{de} = \mp (
\delta_{ec} \eta^{(\pm)i}_{ab} + \delta_{ea} \eta^{(\pm)i}_{bc} -
\delta_{eb} \eta^{(\pm)i}_{ac} ), \\
\label{eta-etabar}
&& \eta^{(\pm)i}_{ab} \eta^{(\mp)j}_{ab}=0, \\
\label{eta^2}
&& \eta^{(\pm)i}_{ac}\eta^{(\pm)j}_{bc} =\delta^{ij}\delta_{ab} +
\varepsilon^{ijk}\eta^{(\pm)k}_{ab}, \\
\label{eta-ex}
&& \eta^{(\pm)i}_{ac}\eta^{(\mp)j}_{bc} =
\eta^{(\pm)i}_{bc}\eta^{(\mp)j}_{ac}, \\
\label{eta-o4-algebra}
&& \varepsilon^{ijk} \eta^{(\pm)j}_{ab} \eta^{(\pm)k}_{cd} =
    \delta_{ac} \eta^{(\pm)i}_{bd} - \delta_{ad} \eta^{(\pm)i}_{bc}
    - \delta_{bc} \eta^{(\pm)i}_{ad} + \delta_{bd} \eta^{(\pm)i}_{ac},
\end{eqnarray}
where $\eta^{(+)i}_{ab} \equiv \eta^i_{ab}$ and $\eta^{(-)i}_{ab} \equiv {\overline
\eta}^i_{ab}$.

If we introduce two families of $4 \times 4$ matrices defined by
\begin{equation} \label{thooft-matrix}
[\tau^i_+]_{ab} \equiv \frac{1}{2} \eta^i_{ab}, \qquad
[\tau^i_-]_{ab} \equiv \frac{1}{2} {\overline \eta}^i_{ab},
\end{equation}
the matrices in (\ref{thooft-matrix}) provide two independent spin $s=\frac{3}{2}$ representations of $su(2)$ Lie algebra.
According to the definition (\ref{tHooft-symbol}), they are explicitly given by
\begin{eqnarray*} \label{t+}
&& \tau^{1}_+ = \frac{1}{2} \begin{pmatrix}
      0 & 0 & 0 & 1 \\
      0 & 0 & 1 & 0 \\
      0 & -1 & 0 & 0 \\
      -1 & 0 & 0 & 0 \\
             \end{pmatrix}, \;\;
  \tau^{2}_+ = \frac{1}{2} \begin{pmatrix}
      0 & 0 & -1 & 0 \\
      0 & 0 & 0 & 1 \\
      1 & 0 & 0 & 0 \\
      0 & -1 & 0 & 0 \\
    \end{pmatrix}, \;\;
  \tau^{3}_+ = \frac{1}{2} \begin{pmatrix}
      0 & 1 & 0 & 0 \\
      -1 & 0 & 0 & 0 \\
      0 & 0 & 0 & 1 \\
      0 & 0 & -1 & 0 \\
    \end{pmatrix}, \\
\label{t-}
&& \tau^{1}_- = \frac{1}{2} \begin{pmatrix}
      0 & 0 & 0 & -1 \\
      0 & 0 & 1 & 0 \\
      0 & -1 & 0 & 0 \\
      1 & 0 & 0 & 0 \\
    \end{pmatrix}, \;\;
  \tau^{2}_- = \frac{1}{2} \begin{pmatrix}
      0 & 0 & -1 & 0 \\
      0 & 0 & 0 & -1 \\
      1 & 0 & 0 & 0 \\
      0 & 1 & 0 & 0 \\
    \end{pmatrix}, \;\;
  \tau^{3}_- = \frac{1}{2} \begin{pmatrix}
      0 & 1 & 0 & 0 \\
      -1 & 0 & 0 & 0 \\
      0 & 0 & 0 & -1 \\
      0 & 0 & 1 & 0 \\
    \end{pmatrix}.
\end{eqnarray*}
The relations in (\ref{eta^2}) and (\ref{eta-ex}) immediately show that
$\tau^i_\pm$ satisfy $su(2)$ Lie algebras, i.e.,
\begin{equation} \label{thooft-su2}
[\tau^i_\pm, \tau^j_\pm] = - \varepsilon^{ijk} \tau^k_\pm,
\qquad [\tau^i_\pm, \tau^j_\mp] = 0.
\end{equation}

\section{Yang-Mills instantons from Einstein manifolds}

In this appendix, we show in detail how to construct the explicit solution of Yang-Mills instantons on
Einstein manifolds with positive curvature, namely $X_4 = \mathbb{S}^4, \; \mathbb{C}P^2, \; \mathbb{S}^2 \times \mathbb{S}^2$
and Page. We will also check whether these manifolds admit a constant instanton density.

Let us briefly recapitulate the method in \cite{yang-col1,oh-yang} and apply it to find the instanton solution
on a general Einstein manifold $X_4$. Consider an Einstein manifold $X_4$ which may be one of $\mathbb{S}^4, \, \mathbb{C}P^2, \,
\mathbb{S}^2 \times \mathbb{S}^2$ and Page. The metric on $X_4$ takes the form
\begin{equation}\label{4-metric}
  ds_4^2 = h_{\alpha\beta} (y) dy^\alpha dy^\beta = e^a \otimes e^a.
\end{equation}
Using the metric, one can calculate the spin connection $\omega_{ab} = \omega_{ab \alpha} dy^\alpha$
and curvature tensor $R_{ab} = \frac{1}{2} R_{ab \alpha\beta} dy^\alpha \wedge dy^\beta$
by solving the structure equations \cite{besse,egh-report}
\begin{eqnarray} \label{t-free}
  &&  T^a = de^a + \omega_{ab} \wedge e^b =0, \\
  \label{curv-eq}
  && R_{ab} = d \omega_{ab} + \omega_{ac} \wedge \omega_{cb}.
\end{eqnarray}
The underlying idea is that Einstein gravity can be formulated as a gauge theory of the Lorentz group,
where spin connections play a role of gauge fields and Riemann curvature tensors correspond to their field
strengths. Another important point is that Riemann curvature tensors, denoted as $R_{ab}$, are $so(4)$-valued
two-forms in $\Omega^2 (X_4) = \Lambda^2 T^* X_4$.
These facts are combined with the well-known theorems \cite{don-kro}:

{\bf Self-duality}. On an orientable Riemannian four-manifold, the 2-forms decompose
into the vector spaces of self-dual and anti-self-dual 2-forms,
\begin{equation}\label{sd-asd}
  \Omega^2 = \Omega_+^2 \oplus \Omega_-^2
\end{equation}
defined by the $\pm 1$ eigenspaces of the Hodge star operator $*:  \Omega^2 \to \Omega^2$.

{\bf Lie group isomorphism}. There is a global isomorphism for the four-dimensional Lorentz group,
i.e., $SO(4)= SU(2)_+ \otimes SU(2)_-/\mathbb{Z}_2$. It also leads to the splitting of the Lie algebra
\begin{equation}\label{lie-iso}
  so(4) = su(2)_+ \oplus su(2)_-.
\end{equation}

\noindent Indeed, these two decompositions are deeply related to each other due to the canonical vector space isomorphism
between the Clifford algebra $\mathbb{C}l(4)$ in four dimensions and the exterior algebra
$\Omega^* M = \bigoplus_{k=0}^4 \Lambda^k T^* M$ over a four-dimensional Riemannian manifold $M$.

We can apply these two decompositions to the spin connection and curvature tensor.
First let us apply the decomposition \eq{lie-iso} to spin connections:
\begin{equation}\label{ga-split}
\omega_{ab} = A^{(+)i} \eta^i_{ab}  +  A^{(-)i}\overline{\eta}^i_{ab}
\end{equation}
where $ \eta^i_{ab}$ and $\overline{\eta}^i_{ab}$ are the 't Hooft symbols satisfying the self-duality relation
\begin{equation}\label{eta-duality}
  \eta^i_{ab} = \frac{1}{2} \varepsilon_{abcd} \eta^i_{cd},
  \qquad  \overline{\eta}^i_{ab} = - \frac{1}{2} \varepsilon_{abcd} \overline{\eta}^i_{cd}.
\end{equation}
See appendix A for the algebraic identities for the 't Hooft symbols.
Thus the spin connections are split into a pair of $SU(2)_+$ and $SU(2)_-$ gauge fields.
Note that the index $i=(1,2,3)$ refers to the $su(2)_\pm$ Lie algebra index.
Accordingly the Riemann curvature tensors $R_{ab}$ are also decomposed into a pair of $SU(2)_+$
and $SU(2)_-$ field strengths:
\begin{equation}\label{gf-split}
R_{ab} = F^{(+)i} \eta^i_{ab}  +  F^{(-)i}\overline{\eta}^i_{ab},
\end{equation}
where $F^{(\pm)i} = \frac{1}{2} F^{(\pm)i}_{ab} e^a \wedge e^b = dA^{(\pm)i} - \varepsilon^{ijk} A^{(\pm)j} \wedge A^{(\pm)k}$.
The second decomposition \eq{sd-asd} is that the six-dimensional vector space of two-forms splits canonically into the sum
of three-dimensional vector spaces of self-dual and anti-self-dual two forms:
\begin{equation} \label{sdf+-}
  F^{(+)i}_{ab} = f_{(++)}^{ij}\eta^j_{ab} + f_{(+-)}^{ij}\overline{\eta}^j_{ab}, \qquad
  F^{(-)i}_{ab} = f_{(-+)}^{ij}\eta^j_{ab} + f_{(--)}^{ij}\overline{\eta}^j_{ab},
\end{equation}
where each component satisfies the self-duality equation according to Eq. \eq{eta-duality}.
Recently, the decomposition \eq{sdf+-} has been used significantly in \cite{chy-grg} to study the scalar invariants
of the curvature tensor.
One can show \cite{oh-yang} that, if $X_4$ is an Einstein manifold obeying the equation $R_{\alpha\beta} = \lambda h_{\alpha\beta}$
with $\lambda$ a cosmological constant,
\begin{equation}\label{eins-cond}
f^{ij}_{(+-)} = 0 = f^{ij}_{(-+)}.
\end{equation}
That is, $SU(2)_+$ and $SU(2)_-$ field strengths in the Riemann curvature tensor \eq{sdf+-} satisfy
the self-duality equations
\begin{equation} \label{hodge-dual}
 F^{(\pm) i}_{ab} = \pm \frac{1}{2} \varepsilon_{abcd} F^{(\pm) i}_{cd}.
\end{equation}
Using the Riemannian metric over $X_4$ in Eq. \eq{4-metric}, the self-duality equations \eq{hodge-dual} can be expressed
in a curved frame as
\begin{equation}\label{sdasd-eq}
  F^{(\pm) i}_{\alpha\beta} = \pm \frac{1}{2} \frac{\varepsilon^{\xi\eta \gamma\delta}}{\sqrt{h}}
  h_{\alpha\xi} h_{\beta\eta} F^{(\pm) i}_{\gamma\delta},
\end{equation}
which is exactly the same as the instanton equation \eq{self-dual-eq}.

\subsection{$\mathbb{S}^4$}

The de Sitter metric on $X_4 = \mathbb{S}^4$ with radius $R = \frac{\zeta_c}{2}$ is
\begin{eqnarray}\label{s4-metric}
  ds^2 &=& h_{\alpha\beta} dy^\alpha dy^\beta = \frac{\zeta_c^4}{(r^2 + \zeta_c^2)^2} dy^\alpha dy^\beta \xx
       &=& \frac{1}{(1+ (\frac{r}{2R})^2 )^2} \left(dr^2 + r^2 \sigma^2_i \right) = e^a \otimes e^a.
\end{eqnarray}
The orthonormal vierbeins $e^a \; (a=1,2,3,4)$ in the metric \eq{s4-metric} are defined by
\begin{equation}\label{vierbein}
 e^a = \left(1+ \Big(\frac{r}{2R} \Big)^2 \right)^{-1} \left\{r \sigma_x, r \sigma_y, r \sigma_z, dr \right\},
\end{equation}
where $\sigma^i \; (i=1,2,3)$ are left-invariant 1-forms on the group manifold $SU(2) \cong \mathbb{S}^3$ and satisfy
the exterior algebra
\begin{equation}\label{esq-s3}
  d\sigma^i + \varepsilon^{ijk} \sigma^j \wedge \sigma^k = 0.
\end{equation}
It is straightforward to calculate the spin connection by solving the torsion-free condition \eq{t-free}.
They read
\begin{equation} \label{spin-s4}
\omega_{ij} = - \varepsilon_{ijk} \sigma^k, \qquad \omega_{i4} = \left(1-\frac{r f'}{f} \right) \sigma^i
\end{equation}
where $f(r) = 1+ \Big(\frac{r}{2R} \Big)^2$ and $f' = \frac{df}{dr}$.
Using the definition \eq{ga-split}, the corresponding $SU(2)$ gauge fields can easily be read off from
the spin connections and they are given by
\begin{eqnarray}\label{su2-gauge}
  A^{(+)i} &=& \frac{1}{4} \eta^i_{ab} \omega_{ab}  \hspace{4cm}  A^{(-)i} = \frac{1}{4} \overline{\eta}^i_{ab} \omega_{ab} \xx
  &=& - \frac{rf'}{2f} \sigma^i = \left( -1 + \frac{1}{f}\right) \sigma^i,
  \hspace{1.9cm}   = - \left( 1- \frac{rf'}{2f}\right) \sigma^i = - \frac{1}{f} \sigma^i,
\end{eqnarray}
where we have used the explicit representation of the 't Hooft symbols.
The $SU(2)$ field strengths are determined by the gauge fields \eq{su2-gauge} as
\begin{eqnarray} \label{su2-field+}
  F^{(+)i} &=& - \frac{1}{2R^2} e^4 \wedge e^i + \frac{1}{4R^2} \varepsilon^{ijk} e^j \wedge e^k, \\
  \label{su2-field-}
  F^{(-)i} &=& \frac{1}{2R^2} e^4 \wedge e^i + \frac{1}{4R^2} \varepsilon^{ijk} e^j \wedge e^k.
\end{eqnarray}
It is straightforward to check the self-duality \eq{hodge-dual}.

It may be instructive to point out that the $SU(2)$ gauge fields in \eq{su2-gauge} coincide with
the instanton solution on $\mathbb{R}^4$. (See Remark 2, p. 296, in \cite{egh-report}.)
Using the coordinates on $\mathbb{R}^4$,
the left-invariant 1-forms $\sigma^i$ on $\mathbb{S}^3$ are represented by \cite{lry-prd2013}
$$ \sigma^i = - \frac{1}{r^2} \eta^i_{\alpha\beta} y^\alpha dy^\beta $$
and then the gauge fields $A^{(+)i}$ in \eq{su2-gauge}, for example, are equal to
\begin{equation}\label{gauge-r4}
 A^{(+)i} = \frac{1}{r^2 + \zeta_c^2} \eta^i_{\alpha\beta} y^\alpha dy^\beta
\end{equation}
where the diameter $2R$ of $\mathbb{S}^4$ is identified with the instanton size $\zeta_c$.
These gauge fields precisely correspond to the gauge fields of a self-dual $SU(2)$ instanton on $\mathbb{R}^4$ \cite{bpst,rajaraman}.
It is not merely a coincidence. The self-duality equation \eq{self-dual-eq} in four dimensions is conformally
invariant, extending to the conformal compactification $\mathbb{S}^4 = \mathbb{R}^4 \cup \{\infty\}$.
Consequently, the solution on $\mathbb{S}^4$ is the same as the finite-action solution on $\mathbb{R}^4$.
This is also related to the fact that the single instanton bundle is the Hopf fibration of $\mathbb{S}^7$ \cite{egh-report}.

For the instanton field strengths in Eqs. \eq{su2-field+} and \eq{su2-field-}, we get
\begin{equation}\label{inst-density}
  F^{(\pm)i} \wedge F^{(\pm)i} = \pm \frac{3}{2R^4} \sqrt{h} d^4 y.
\end{equation}
This result exactly coincides with Eq. (3.19) in \cite{kky} (see footnote \ref{diff-ggt}).
So we confirm that the Yang-Mills instanton obtained from the spin connection of $\mathbb{S}^4$ admits a constant density.

\subsection{$\mathbb{C}P^2$}

The metric on $\mathbb{C}P^2$ takes the form
\begin{eqnarray}\label{cp2-metric}
  ds^2 &=& h_{\alpha\beta} dy^\alpha dy^\beta = \frac{1}{f} (r^2 \sigma_1^2 + r^2 \sigma_2^2)
       + \frac{1}{f^2} (r^2 \sigma^2_3 + dr^2) = e^a \otimes e^a
\end{eqnarray}
where
$$ f(r) = 1 + \frac{\lambda r^2}{6} $$
and the orthonormal vierbeins are defined by
\begin{equation*}
  e^a = \left\{\frac{r \sigma_x}{\sqrt{f}}, \frac{r \sigma_y}{\sqrt{f}}, \frac{r \sigma_z}{f}, \frac{dr}{f} \right\}.
\end{equation*}
The spin connections read as
\begin{eqnarray*} \label{spin-cp2}
&& \omega_{12} = (f^{-1} -2) \sigma^3 = - \frac{1 + \frac{\lambda r^2}{3}}{1 + \frac{\lambda r^2}{6}} \sigma^3,
\quad \omega_{13} = \frac{1}{\sqrt{f}} \sigma^2, \quad \omega_{23} = - \frac{1}{\sqrt{f}} \sigma^1, \\
&& \omega_{14} =  \frac{1}{\sqrt{f}} \sigma^1,
\quad \omega_{24} = \frac{1}{\sqrt{f}} \sigma^2, \quad \omega_{34} = \left( 1 - \frac{r f'}{f} \right) \sigma^3
= \frac{1 - \frac{\lambda r^2}{6}}{1 + \frac{\lambda r^2}{6}} \sigma^3.
\end{eqnarray*}
The corresponding $SU(2)$ gauge fields can be read off from the spin connections as
\begin{eqnarray}\label{cp2-gauge}
&&  A^{(+)1} = A^{(+)2} = 0, \qquad A^{(+)3} = - \frac{1}{4} \lambda r e^3, \xx
&&  A^{(-)1} = - \frac{1}{r} e^1, \quad  A^{(-)2} = - \frac{1}{r} e^2,
\quad A^{(-)3} = - \frac{1+ f}{2r} e^3.
\end{eqnarray}
The field strengths are determined from the above gauge fields as
\begin{eqnarray} \label{cp2-field+}
 &&  F^{(+)1} = F^{(+)2} = 0,  \quad F^{(+)3} = d A^{(+)3} = \frac{\lambda}{2} (e^1 \wedge e^2 + e^3 \wedge e^4), \\
  \label{cp2-field-}
  && F^{(-)i} = \frac{\lambda}{6} \left( \frac{1}{2} \varepsilon^{ijk} e^j \wedge e^k - e^i \wedge e^4 \right).
\end{eqnarray}
In particular, the self-dual sector is an Abelian gauge field.
The self-duality \eq{hodge-dual} is manifest.

In order to check whether the instanton density, $\rho^{(\pm)}_n = - \frac{1}{4} \mathrm{Tr} 
F^{(\pm)}_{\alpha\beta} F^{(\pm) \alpha\beta}$, is constant, let us calculate the quantity
\begin{equation}\label{app-density}
  F^{(\pm)i} \wedge F^{(\pm)i} = \pm 2 \rho^{(\pm)}_n \sqrt{h} d^4 y
\end{equation}
where $\sqrt{h} d^4 y = e^1 \wedge e^2 \wedge e^3 \wedge e^4$ is the volume form on $\mathbb{C}P^2$.
For the instanton solution in Eqs. \eq{cp2-field+} and \eq{cp2-field-}, we get
\begin{equation}\label{cps-density}
  F^{(+)i} \wedge F^{(+)i} = \frac{\lambda^2}{2} \sqrt{h} d^4 y,
  \qquad F^{(-)i} \wedge F^{(-)i} = - \frac{\lambda^2}{6} \sqrt{h} d^4 y.
\end{equation}
So we confirm that the instanton density on $\mathbb{C}P^2$ is also constant.

\subsection{$\mathbb{S}^2 \times \mathbb{S}^2$}

Let us take the metric on $X_4 = \mathbb{S}^2 \times \mathbb{S}^2$ as the form
\begin{eqnarray}\label{s2s2-metric}
  ds^2 &=& h_{\alpha\beta} dy^\alpha dy^\beta = \sum_{\flat=1}^2 R_\flat^2(d\theta_\flat^2 + \sin^2 \theta_\flat d\phi_\flat^2) = e^a \otimes e^a
\end{eqnarray}
where the orthonormal vierbeins are given by
\begin{equation*}
  e^a = \left\{ R_1 d\theta_1, R_1 \sin \theta_1 d\phi_1, R_2 d\theta_2, R_2 \sin \theta_2 d\phi_2 \right\}.
\end{equation*}
The nonzero spin connections are
\begin{equation*}
\omega_{12} = - \frac{\cot \theta_1}{R_1}e^2, \qquad \omega_{34} = - \frac{\cot \theta_2}{R_2} e^4.
\end{equation*}

The corresponding gauge fields consist of two $U(1)$ gauge fields
\begin{eqnarray}\label{u12-gauge}
&& A^{(+)3} = - \frac{\cot \theta_1}{2R_1}e^2 - \frac{\cot \theta_2}{R_2} e^4, \xx
&& A^{(-)3} = -  \frac{\cot \theta_1}{2R_1}e^2 + \frac{\cot \theta_2}{R_2} e^4.
\end{eqnarray}
The field strengths are determined by the above gauge fields as
\begin{eqnarray} \label{s2-field+}
 && F^{(+)3} = d A^{(+)3} = \frac{1}{2} \left(\frac{e^1 \wedge e^2}{R_1^2} + \frac{e^3 \wedge e^4}{R_2^2} \right), \\
 \label{s2-field-}
 && F^{(-)3} = d A^{(-)3} = \frac{1}{2} \left(\frac{e^1 \wedge e^2}{R_1^2} - \frac{e^3 \wedge e^4}{R_2^2} \right).
\end{eqnarray}
The self-duality \eq{hodge-dual} requires that $R_1 = R_2$.
For the above field strengths, we obtain
\begin{equation}\label{s2-density}
  F^{(\pm)3} \wedge F^{(\pm)3} = \pm \frac{1}{2 R_1^2 R_2^2} \sqrt{h} d^4 y.
\end{equation}
Thus we get a constant instanton density even if $R_1 \neq R_2$.
It is possible because $X_4 = \mathbb{S}^2 \times \mathbb{S}^2$ is a product space.

\subsection{Page}

The final example is the Page space $X_4 = \mathbb{C}P^2 \sharp \overline{\mathbb{C}P^2}$
which is an $\mathbb{S}^2$ bundle over $\mathbb{S}^2$ \cite{page-metric}.
It is constructed by attaching the complex projective space and its complex conjugate,
removing a 4-ball from each manifold, and gluing them together
along the resulting $\mathbb{S}^3$ boundaries. The metric on the Page space takes on a complicated form
\begin{eqnarray}\label{page}
  ds^2 &=& \frac{3}{\lambda} \frac{1+\nu^2}{3+\nu^2} \left( U^{-1} d\chi^2
  + \frac{U}{4} \sin^2 \chi (d\psi+ \cos \theta d\phi)^2
  + \frac{1-\nu^2 \cos^2 \chi}{4\nu} (d\theta^2 + \sin^2 \theta d\phi^2) \right), \xx
  &=& e^a \otimes e^a
\end{eqnarray}
where
$$ U(\chi) = 1 - \frac{2 \nu^2 \sin^2 \chi}{(3+ \nu^2)(1-\nu^2 \cos^2 \chi)} $$
and $\nu$ is the positive root of $\nu^4 + 4 \nu^3 - 6\nu^2 + 12\nu - 3=0$ which works out to be $0.2817$.
The orthonormal vierbeins are defined by
\begin{equation*}
 \sqrt{\frac{\lambda}{3} \frac{3+\nu^2}{1+\nu^2}} e^a = \left\{\sqrt{\frac{1-\nu^2 \cos^2 \chi}{4\nu}} d\theta,
 \sqrt{\frac{1-\nu^2 \cos^2 \chi}{4\nu}} \sin \theta d\phi,
 \frac{\sqrt{U}}{2}\sin \chi (d\psi+ \cos \theta d\phi), \frac{d\chi}{\sqrt{U}} \right\}.
\end{equation*}
Here we choose $\lambda = 3 \frac{1+\nu^2}{3+\nu^2}$ for simplicity.
After a little algebra, we get the spin connections
\begin{eqnarray}
&& \omega_{12} = - \sqrt{\frac{4\nu}{1-\nu^2 \cos^2 \chi}} \cot \theta e^2 + f e^3,
\quad \omega_{13} = f e^2, \quad \omega_{23} = - f e^1,   \xx
&& \omega_{14} = g e^1,  \quad  \omega_{24} = g e^2,
\quad    \omega_{34} = \left( \sqrt{U} \cot \chi + \frac{2 (\nu^2 -1)}{U (3+\nu^2) (1-\nu^2 \cos^2 \chi)} g  \right) e^3,
\end{eqnarray}
where
$$ f= \frac{\sqrt{U} \nu}{1-\nu^2 \cos^2 \chi} \sin \chi,  \qquad g =  f \nu \cos \chi.$$

\begin{figure}
\centering
\includegraphics[width=0.4\textwidth]{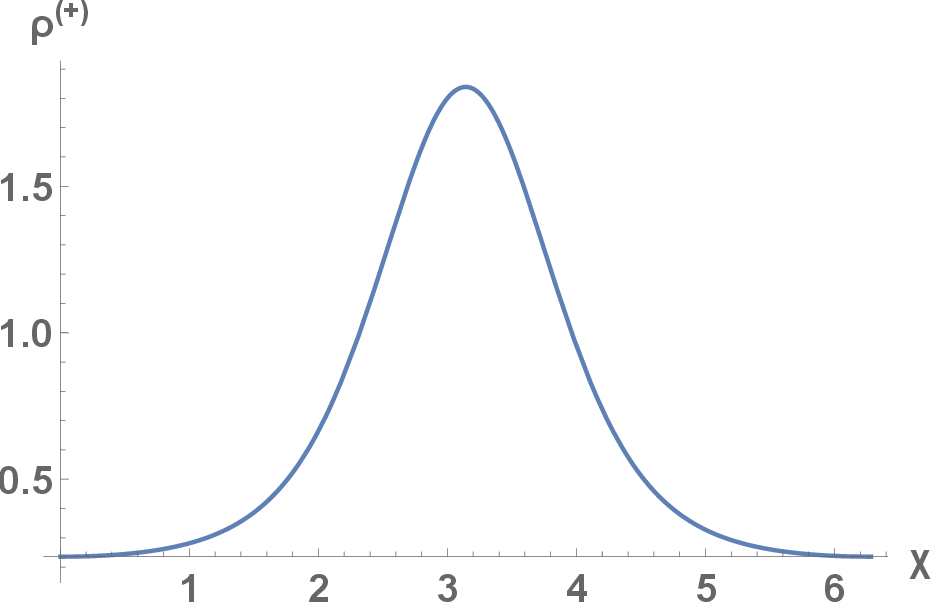}
\includegraphics[width=0.4\textwidth]{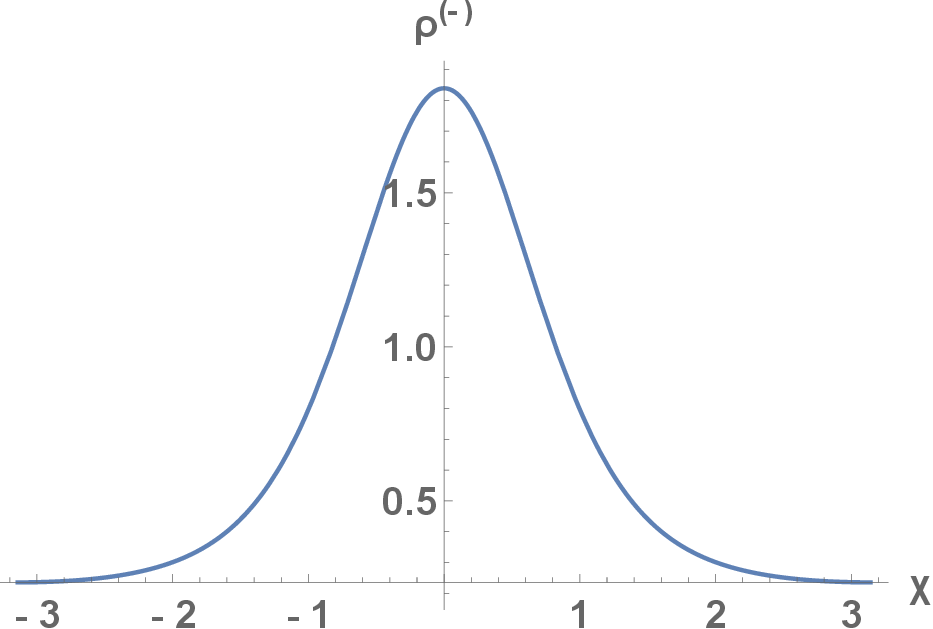}
\caption{\label{inst-den} The instanton densities $\rho^{(\pm)}$
for self-dual and anti-self-dual instantons on $X_4 = $ Page.}
\end{figure}

The corresponding $SU(2)$ gauge fields are given by
\begin{eqnarray*} \label{page-gauge}
  A^{(\pm)1} &=& \frac{1}{2} (-f \pm g) e^1, \qquad A^{(\pm)2} = \frac{1}{2} (-f \pm g) e^2, \\
  A^{(\pm)3} &=& \frac{1}{2} \left( - \sqrt{\frac{4\nu}{1-\nu^2 \cos^2 \chi}} \cot \theta e^2 + f e^3
  \pm \Big( \sqrt{U} \cot \chi + \frac{2 (\nu^2 -1)}{U (3+\nu^2) (1-\nu^2 \cos^2 \chi)} g \Big) e^3 \right).
\end{eqnarray*}
It is straightforward (but a bit tedious) to calculate the $SU(2)$ field strengths:
\begin{eqnarray}\label{page-f}
  F^{(\pm)1} &=& \frac{U}{2} \frac{\nu(1-\nu^2) \cos \chi \mp \nu^2 (\cos^2 \chi - \sin^2 \chi)
  \pm \nu^4 \cos^4 \chi}{(1-\nu^2 \cos^2 \chi)^2} (e^1 \wedge e^4 \pm e^2 \wedge e^3), \xx
  && - \frac{\nu^3 (1-\nu^2) \cos \chi \sin^2 \chi}{(3+\nu^2)(1 - \nu^2 \cos^2 \chi)^3} (1 \mp \nu\cos \chi)
  (e^1 \wedge e^4 \pm e^2 \wedge e^3), \xx
  F^{(\pm)2} &=& \frac{U}{2} \frac{\nu(1-\nu^2) \cos \chi \mp \nu^2 (\cos^2 \chi - \sin^2 \chi)
  \pm \nu^4 \cos^4 \chi}{(1-\nu^2 \cos^2 \chi)^2} (e^2 \wedge e^4 \pm e^3 \wedge e^1) \xx
   && - \frac{\nu^3 (1-\nu^2) \cos \chi \sin^2 \chi}{(3+\nu^2)(1-\nu^2 \cos^2 \chi)^3} (1 \mp \nu\cos \chi)
  (e^2 \wedge e^4 \pm e^3 \wedge e^1), \\
  F^{(\pm)3} &=& - \frac{\nu}{2(1 - \nu^2 \cos^2 \chi)}
  \left( U' \sin \chi + \frac{2(1-\nu^2) U \cos \chi}{1 - \nu^2 \cos^2 \chi} \right) (e^3 \wedge e^4 \pm e^1 \wedge e^2 ) \xx
&& \pm \frac{\nu}{2(1 - \nu^2 \cos^2 \chi)}
  \left( 4-  \frac{\nu U \sin^2 \chi (3+ \nu^2 \cos^2 \chi)}{1 - \nu^2 \cos^2 \chi} \right)
  (e^3 \wedge e^4 \pm e^1 \wedge e^2). \nonumber
\end{eqnarray}
One can show that the Ricci tensor is given by (after recovering the factor $\frac{3}{\lambda} \frac{1+\nu^2}{3+\nu^2}$)
\begin{align}
R^{(0)}_{ab} = \lambda\, \text{diag} \left( c, c, 1, 1 \right),
\end{align}
where $c =\frac{\nu ^4-8 \nu ^3+9 \nu ^2-24 \nu +3 \left(\nu ^2+1\right) \nu ^2 \cos 2 \chi }{3 \nu ^4-3 \nu ^2-6+3
\left(\nu ^2+1\right) \nu ^2 \cos 2 \chi }$. Using the quartic equation $\nu^4 + 4 \nu^3 - 6\nu^2 + 12\nu - 3=0$,
one can show that $c=1$. Therefore the Page space is an Einstein manifold that satisfies the Einstein equation
$R^{(0)}_{\alpha\beta}= \lambda h_{\alpha\beta}$.
Instead of showing the full expressions, we present the plots of the instanton densities in Fig. \ref{inst-den}.
The full expression is highly complex and does not provide any useful insight.
The figure clearly illustrates that Yang-Mills instantons on the Page space exhibit an inhomogeneous density.

\subsection{Squashed $\mathbb{S}^4$}

In this example, we aim to understand the importance of the homogeneity of an internal space $X_4$ in achieving
a constant instanton density on $X_4$. We examine another nonhomogeneous space obtained by deforming a round sphere $\mathbb{S}^4$
whose deformation is parameterized by three axis-length parameters $(r, l, \tilde{l})$.
The metric of squashed 4-sphere is given by $ds^2 = e^a \otimes e^a$ with the vierbeins \cite{squashed-s4}
\begin{equation*}\label{sps4-vierbein}
  \left\{ e^1 = \sin \rho \tilde{\sigma}^1, \; e^2 = \sin \rho \tilde{\sigma}^2, \;
e^3 = \sin \rho \tilde{\sigma}^3 + h d\rho, \; e^4 = g d\rho \right\},
\end{equation*}
where
\begin{eqnarray*}
  f &\equiv & \sqrt{l^2 \sin^2 \theta + \tilde{l}^2 \cos^2 \theta}, \\
  g &\equiv & \sqrt{r^2 \sin^2 \rho + l^2 \tilde{l}^2 f^{-2} \cos^2 \rho}, \\
  h &\equiv & \frac{\tilde{l}^2 -l^2}{f} \cos \rho \sin \theta \cos \theta,
\end{eqnarray*}
and $\tilde{\sigma}^i \, (i=1,2,3)$ are dreibeins of the three-dimensional ellipsoid
given in polar coordinates $(\varphi, \chi, \theta)$ by
\begin{equation*}
  \tilde{\sigma}^1 = l \cos \theta d \varphi, \;
  \tilde{\sigma}^2 = \tilde{l} \sin \theta d \chi, \; \tilde{\sigma}^3 = f d \theta.
\end{equation*}

It is straightforward to calculate the $SU(2)$ gauge fields
\begin{eqnarray} \label{sqs4-gauge}
  A^{(\pm)1} &=& \pm \frac{\tilde{l}^2}{2 gf^2} \cot \rho e^1 + \frac{\cot \theta}{2 f \sin \rho}e^2, \xx
  A^{(\pm)2} &=& \pm \frac{l^2}{2 gf^2} \cot \rho e^2 + \frac{\tan \theta}{2 f \sin \rho}e^1, \\
  A^{(\pm)3} &=& \pm \frac{l^2 \tilde{l}^2}{2 gf^4} \cot \rho \left( e^3 - \frac{h}{g} e^4 \right), \nonumber
\end{eqnarray}
and their field strengths
\begin{eqnarray}\label{sqs4-f}
  F^{(\pm)1} &=& \pm \frac{\tilde{l}^2}{2\sin \rho} d\psi \wedge e^1
  \pm \frac{\tilde{l}^2 (\tilde{l}^2 - l^2) \psi \sin 2 \theta}{4 f^3 \sin^2 \rho} e^1 \wedge (e^3 - g^{-1} h e^4) \xx
  && + \frac{l^2 r^2}{2 f^4 g^2}  e^2 \wedge (e^3 - g^{-1} h e^4), \xx
  F^{(\pm)2} &=& \pm \frac{l^2}{2 \sin \rho} d\psi \wedge e^2
  \pm \frac{l^2 (\tilde{l}^2 - l^2) \psi \sin 2 \theta}{4 f^3 \sin^2 \rho} e^2 \wedge (e^3 - g^{-1} h e^4) \\
  && - \frac{\tilde{l}^2 r^2}{2 f^4 g^2} e^1 \wedge (e^3 - g^{-1} h e^4), \xx
  F^{(\pm)3} &=& \pm \frac{l^2 \tilde{l}^2}{2 f^2 \sin \rho} d\psi \wedge (e^3 - g^{-1} h e^4)
  + \frac{r^2}{2f^2 g^2} e^1 \wedge e^2, \nonumber
\end{eqnarray}
where $\psi \equiv \frac{\cos \rho}{ g f^2}$ and $d \psi = - \frac{r^2 \sin \rho}{g^4 f^2} e^4 + \frac{h}{g^3 f^6 \sin \rho}
(2 g^2 f^2 - l^2 \tilde{l}^2 \cos^2 \rho ) (e^3 - g^{-1} h e^4)$.

The field strengths in Eq. \eq{sqs4-f} are neither self-dual nor anti-self-dual,
so the squashed four-sphere is not an Einstein manifold. According to the result \eq{eins-cond},
$SU(2)$ gauge fields constructed from a four-manifold $X_4$ become self-dual or anti-self-dual only
if $X_4$ is Einstein, i.e., obeying the vacuum Einstein equation, $R_{\alpha\beta} = \lambda h_{\alpha\beta}$ \cite{oh-yang}.
As a result, $SU(2)$ field strengths in Eq. \eq{sqs4-f} do not satisfy the self-duality \eq{hodge-dual}.
Therefore the squashed four-sphere would be relevant as an internal manifold only when matters are coupled with gravity.
It is straightforward to see how the instanton density behaves for the squashed four-sphere.
The instanton density for the $SU(2)$ field strengths in \eq{sqs4-f} exhibits a nontrivial distribution as expected:
\begin{eqnarray} \label{inst-dens}
F^{(\pm)i} \wedge F^{(\pm)i} = \pm \frac{3}{2} \frac{l^2 \tilde{l}^2 r^4}{f^6 g^6} \sqrt{h} d^4 y,
\end{eqnarray}
where $e^1 \wedge e^2 \wedge e^3 \wedge e^4 = \sqrt{h} d^4 y$ is the volume form on the squashed four-sphere.
When $l = \tilde{l} = r$, it becomes the round four-sphere $\mathbb{S}^4$ which is an Einstein manifold.
In this case, the $SU(2)$ gauge fields in \eq{sqs4-gauge} become instanton connections and
the instanton density \eq{inst-dens} becomes constant, i.e., $F^{(\pm)i} \wedge F^{(\pm)i} = \pm \frac{3}{2 r^4} \sqrt{h} d^4 y$
which is consistent with \eq{inst-density} in the unit $R=r$.
But the squashed $\mathbb{S}^4$ has a non-uniform distribution.
This result supports our claim that the instanton density on a nonhomogeneous manifold is not uniform in general.

\end{document}